\documentclass[danish,a4paper,10pt,twocolumn,amsmath,amssymb,floatfix,superscriptaddress,aps,pra,showpacs,longbibliography]{revtex4-1}
\usepackage{graphicx}
\usepackage[english]{babel}
\usepackage{bm}
\usepackage{dcolumn}
\allowdisplaybreaks[4]

\newcommand{\fref}[1]{Fig.~\ref{#1}}

\newcommand{\be}{\begin{equation}}
\newcommand{\ee}{\end{equation}}
\newcommand{\ba}{\begin{eqnarray}}
\newcommand{\ea}{\end{eqnarray}}
\newcommand{\bs}{\begin{subequations}}
\newcommand{\es}{\end{subequations}}
\newcommand{\bw}{\begin{widetext}}
\newcommand{\ew}{\end{widetext}}

\usepackage{color}
\definecolor{red}{rgb}{1,0,0}
\definecolor{blue}{rgb}{0,0,1}
\usepackage{MnSymbol}

\begin{document}

\title{Nuclear-Motion Effects in Attosecond Transient Absorption Spectroscopy of Molecules}

\author{Jens~E.~B\ae kh\o j}
\affiliation{Department of Physics and Astronomy, Aarhus University, 8000 Aarhus C, Denmark}

\author{Lun~Yue}
\affiliation{Department of Physics and Astronomy, Aarhus University, 8000 Aarhus C, Denmark}

\author{Lars~Bojer~Madsen}
\affiliation{Department of Physics and Astronomy, Aarhus University, 8000 Aarhus C, Denmark}

\date \today
\begin{abstract}
We investigate the characteristic effects of nuclear motion on attosecond transient absorption spectra in molecules by calculating the spectrum for different model systems. Two models of the hydrogen molecular ion are considered: one where the internuclear separation  is fixed, and one where the nuclei are free to vibrate. The spectra for the fixed nuclei model are similar to atomic spectra reported elsewhere, while the spectra obtained in the model including nuclear motion are very different and dominated by extremely broad absorption features. These broad absorption features are analyzed and their relation to molecular dissociation investigated. 
The study of the hydrogen molecular ion validates an approach based on the Born-Oppenheimer approximation and a finite electronic basis. 
This latter approach is then used to study the three-dimensional hydrogen molecule including nuclear vibration.
The spectrum obtained from H$_2$ is compared to the result of a fixed-nuclei calculation.
In the attosecond transient absorption spectra of H$_2$ including nuclear motion we find a rich absorption structure corresponding to population of different vibrational states in the molecule, while the fixed-nuclei spectra again are very similar to atomic spectra. We find that light-induced structures at well-defined energies reported in atomic systems are also present in our fixed nuclei molecular spectra, but suppressed in the ${\text{H}_2}^+$ and H$_2$ spectra with moving nuclei. We show that the signatures of light-induced structures are closely related to the nuclear dynamics of the system through the shapes and relative arrangement of the Born-Oppenheimer potential energy curves.
\end{abstract}

\pacs{33.80.Wz, 33.20.Tp}


\maketitle

\section{\label{introduction} Introduction}
Recent progress in laser technology has made it possible to produce isolated attosecond pulses \cite{hentschel2001attosecond}, and with these electron dynamics can be followed on its natural time scale \cite{krausz2009attosecond}. 
One of the promising methods to follow sub-femtosecond dynamics in electronic systems is attosecond transient absorption spectroscopy (ATAS), which offers an all-optical approach to light-matter interactions, with the high time resolution of the attosecond pulse and the high energy resolution characteristic of absorption spectroscopy.
Since 2010, ATAS has proven itself as a very efficient tool in studying electron dynamics in atoms.
By following the movement of electron distributions in a time integrated manner, ATAS experiments have been used to study a number of dynamical processes.
Examples hereof are the observation of valence electron motion in krypton ions \cite{goulielmakis2010real}, autoionization in argon \cite{wang2010attosecond}, the observation of AC Stark shifts of excited states in helium \cite{chini2012subcycle} and  krypton \cite{wirth2011synthesized}, and very recently observation and control of the dynamics in a two-electron wave packet of helium \cite{ott2014reconstruction}.
Common for these experiments is that a near-infrared (NIR) or infrared (IR) pulse dresses the system, while the modification of the spectrum of a much shorter extreme ultraviolet (XUV) attosecond pulse is measured. There is some ambiguity in the literature about which pulse should be refereed to as the pump (probe) pulse in the pump-probe setup of ATAS. 
We circumvent this problem by referring to the two pulses simply as the 'NIR pulse' and the 'XUV pulse'. 
A variable delay between the XUV and NIR pulses makes it possible to introduce an indirect time-resolution in time-integrated ATAS spectra, and in this way follow the evolution of the NIR field-dressed system.
Theoretical investigations of ATAS have focused on explaining some of the typical features in the spectra such as Autler-Townes splittings of absorption lines \cite{pfeiffer2012transmission,wu2013time}, light-induced structures (LIS) \citep{chen2012light}, and different interference phenomena \cite{chen2013quantum,chini2014resonance}. 
Often simplified few-level models can capture the essential physics of a process and one can benefit from the fact that the moderate intensity of the XUV pulse places its interaction with the quantum system in the perturbative regime \cite{chini2012subcycle}.
Finally an increasing number of direct calculations of the delay-dependent ATAS intensity spectrum has been carried out by solving the time-dependent Schr\"odinger equation (TDSE) for a variety of systems. Examples hereof are given in Refs.~\cite{ott2014reconstruction, chen2012light} where the ATAS intensity spectra were found by solving the TDSE for helium. In the latter case the single-active-electron approximation was used.

Very recently ATAS experiments have been carried out on nitrogen \cite{sansone2013attosecond} and hydrogen \cite{cheng2014attosecond} molecules. Neglecting the slow molecular rotations, the main difference between atoms and di-atomic molecules is that molecular systems vibrate in the internuclear separation.
In view of the emerging experimental interest in ATAS of molecules it is important to characterize the main effects associated with the nuclear vibration. Such a characterization is the aim of the present work.
First we investigate one-dimensional (1D) ${\text{H}_2}^+$ by solving the TDSE exactly within the limitations of our two models: one where the nuclear distance is fixed and one where the nuclear degree of freedom is included. 
We note in passing that there is a long history in strong-field physics for using co-linear ${\text{H}_2}^+$ as a generic system to highlight molecular effects \cite{Kulander96,ver2003time,Weixing01}. 
The TDSE calculations in ${\text{H}_2}^+$ show that the dissociative nature of the excited electronic states in the moving-nuclei model leads to major differences in the ATAS spectrum when compared to the fixed-nuclei case. 
The fixed-nuclei spectra are very similar to atomic spectra. In both cases the spectra are dominated by narrow, well-separated, absorption features centered around energies corresponding to bound states of the system. In moving-nuclei spectra, on the other hand, the absorption features are much broader and not necessarily well-separated.
In the analysis of these broad moving-nuclei induced absorption lines we establish a method to determine the dissociation time of nuclear wave packets evolving on certain excited BO surfaces. This method does not depend on a pump-probe setup, and information about the dissociation time can be obtained from single-pulse experiments.
We also find that the ATAS spectra obtained from TDSE calculations can be reproduced accurately using an approach based on the Born-Oppenheimer (BO) approximation and an expansion in a finite electronic basis. 
This approach is then used to calculate the ATAS spectrum of three-dimensional (3D) $\text{H}_2$ including nuclear motion. Compared to ${\text{H}_2}^+$, H$_2$ represent a case where  the lowest excited electronic states are binding.
In the ATAS spectrum of ${\text{H}_2}$ we see a rich structure associated with population transfer to vibrational states of excited BO surfaces. In the spectrum of ${\text{H}_2}$ we recognize some of the characteristics found in fixed nuclei ${\text{H}_2}^+$ and atomic spectra, however, other effects such as Autler-Townes splittings and LIS at well-defined energies are absent in the moving nuclei spectrum. Our analysis show that the suppression of such LIS in moving nuclei systems are directly related to the molecular dynamics of the system, and that LIS in molecules in general are extremely dependent on the population of the involved vibrational states and thereby on the BO curves of the involved electronic states.

The paper is organized as follows. In Sec.~\ref{theory} we describe a theory for calculating the ATAS intensity spectrum. We use the single-system response model of Ref.~\cite{baggesen2011dipole} (see also Ref.~\cite{baggesen2012theory}). Within the approximations given in Sec.~\ref{theory}, the ATAS intensity spectrum is readily determined if the dipole moment of the system is known for all times. 
In Sec.~\ref{theory} we also establish the relationship between our method and other methods often used to calculate the ATAS spectrum \cite{ott2014reconstruction,gaarde2011transient,santra2011theory}.
In Sec.~\ref{results} we present calculated ATAS spectra for our two models of ${\text{H}_2}^+$, and discuss the atomic-like characteristics of the model with fixed nuclei. A three-level model description capturing some of the interference features is discussed in Appendix~\ref{app}.
In the model of ${\text{H}_2}^+$ where the nuclei are allowed to move, we find that the absorption lines in the ATAS spectrum become very broad. In Sec.~\ref{discussion} we set up a model to explain these broad absorption lines, and use it to closer analyze the dissociation process of excited electronic states. 
In Sec.~\ref{H2} we present the ATAS spectrum for 3D $\text{H}_2$ for fixed and moving nuclei. 
The fixed nuclei spectrum is very similar to the atomic-like, fixed nuclei spectrum of ${\text{H}_2}^+$.
However, as a result of population of vibrational states, corresponding to different excited electronic states, the moving nuclei ATAS spectrum of $\text{H}_2$ is very complex, and we will focus on the overall structure, interference phenomena in the absorption lines and LIS.
Finally Sec.~\ref{conclusion} concludes the paper. 
Atomic units ($\hbar=e=m_e=a_0=1$) are used throughout unless indicated otherwise.

\section{\label{theory} Theory}
The description of a laser pulse propagating though a medium in principle requires a full quantum mechanical description of both the field and the medium. In this work, however, all pulses considered are far too intense to show any non-classical behavior, and the fields are therefore treated classically throughout. 
Even in this semi-classical limit the full description of ATAS requires that we solve the TDSE and Maxwell's equations simultaneously, which is computationally very demanding \cite{gaarde2011transient}.
In Ref.~\cite{chen2012light}, it is argued that the ATAS intensity spectrum of helium found using a single-atom response model, does not vary qualitatively from the spectrum found using a description including light propagation if the target density is not too high.
A similar conclusion can be drawn from the work presented in Ref.~\cite{ott2014reconstruction}.
Since our field parameters are very similar to the ones used in Ref.~\cite{chen2012light}, and our main focus is a qualitative description of characteristic molecular effects in ATAS, we expect the single-system response model to be sufficiently  accurate for our purpose, and we will use it throughout the paper. To calculate how the XUV pulse is modified by propagation in ATAS experiments, we follow the approach of Ref.~\cite{baggesen2012theory}.

From Maxwell's equations we obtain the wave equation for the electromagnetic field $\textbf{E}(\textbf{r},t)$
\begin{align}
\left( \nabla^2-\frac{1}{c^2}\frac{\partial^2}{\partial t^2} \right)\textbf{E}(\textbf{r},t)=\frac{4\pi}{c^2}\frac{\partial^2}{\partial t^2} \textbf{P}(\textbf{r},t),
\label{wave_gen}
\end{align}
where $c \simeq 137$ is the speed of light in vacuum and $\textbf{P}(\textbf{r},t)$ is the time-dependent polarization of the target defined as the dipole moment density in the target region.
We consider fields linearly polarized along the $z$-direction, and propagating in the positive $x$-direction.
We apply the single-system response model, thereby reducing the interaction between field and matter to a single point in space and for the $z$-components of $\textbf{E}(\textbf{r},t)$ and $\textbf{P}(\textbf{r},t)$ we obtain
\begin{align}
\left( \frac{\partial^2}{\partial x^2}-\frac{1}{c^2}\frac{\partial^2}{\partial t^2} \right)E(x,t)=\frac{4\pi}{c^2}\frac{\partial^2}{\partial t^2} \delta(x) P(t).
\label{wave_eq}
\end{align}
The polarization is given in terms of the $z$-component of the single-atom or single-molecule time-dependent dipole moment
\begin{align}
P(t)=n \langle d(t) \rangle,
\label{Pol}
\end{align}
where $n$ is the density of atoms or molecules in the sample and $\langle d(t) \rangle$ is the expectation value of the dipole operator $\langle d(t) \rangle=\langle \Psi(t) \vert d \vert \Psi(t) \rangle$, with $\Psi(t)$ the solution of the TDSE. The classical field in Eq.~\eqref{wave_eq} is therefore linked to the quantum evolution of the system through $P(t)$.
We now split the laser field in an incoming field and a field generated by the light-matter interaction. We assume that the incoming field propagates unchanged for all times
\begin{align}
E(x,t)=E_\text{in} \left(t-\frac{x}{c} \right) + E_\text{gen}(x,t).
\label{field_tot}
\end{align}
All modifications of the total field are thus included in the generated field $E_\text{gen}(x,t)$.
For the electric field of Eq.~\eqref{field_tot}, the wave equation \eqref{wave_eq} can be solved analytically. The solution for the generated field is \cite{baggesen2012theory}
\begin{align}
E_\text{gen}(x,t)= &-\frac{2\pi}{c} \left[ \theta(x) \frac{\partial}{\partial t} P \left(t-\frac{x}{c} \right) + \theta(-x) \frac{\partial}{\partial t} P \left(t+\frac{x}{c} \right) \right],
\label{field_gen}
\end{align} 
where $\theta(x)$ is the heaviside function. From Eq.~\eqref{field_gen} we see that the generated field has two contributions propagating in opposite directions.

In the observation point $x$, the intensity spectrum corresponding to the electromagnetic field $E(x,t)$ is expressed as
\begin{align}
S(\omega) &\propto \left\vert \frac{1}{2\pi} \int_{-\infty}^{\infty} E(x,t) \exp(i\omega t) dt \right\vert^2 \nonumber \\
&=\left\vert E_\text{in}(\omega) \right\vert^2 + 2\text{Re}\left[ E_\text{in}^*(\omega) E_\text{gen}(\omega) \right] + \left\vert E_\text{gen}(\omega) \right\vert^2,
\label{spec_tot}
\end{align}
where $E_\text{in}(\omega)$ and $E_\text{gen}(\omega)$ are the Fourier components
\begin{align}
E_\text{in}(\omega)&=\frac{1}{2\pi}\int_{-\infty}^{\infty} E_\text{in} \left( t-\frac{x}{c} \right) \exp(i\omega t) dt
\label{F_E_in}
\end{align}
\vspace{-5ex}
\begin{subequations}
\begin{align}
E_\text{gen}(\omega)&=\frac{1}{2\pi}\int_{-\infty}^{\infty} E_\text{gen} \left( x,t \right) \exp(i\omega t) dt \label{pre_F_E_gen} \\
&=\frac{\omega i n}{c} \int_{-\infty}^{\infty} \langle d(t) \rangle \exp(i\omega t) dt.
\label{F_E_gen}
\end{align}
\end{subequations}
Equation \eqref{F_E_gen} is obtained from Eq. \eqref{pre_F_E_gen} by partial integration, only retaining the solution propagating in the positive $x$-direction, and finally Eq.~\eqref{Pol} is used.
We now consider the three terms of Eq.~\eqref{spec_tot}. 
The first term contains only information about the incoming field, and is therefore not of any interest, since we want to investigate the modification of the XUV pulse induced by the target. The two remaining terms contain information about this modification. However, since $E_\text{gen} \ll E_\text{in}$, the second term dominates over the third. Henceforth we therefore refer to
\begin{align}
\tilde{S}(\omega,\tau)=2\text{Re}\left[ E_\text{in}^*(\omega,\tau) E_\text{gen}(\omega,\tau) \right],
\label{response}
\end{align}
as the ATAS spectrum. As seen from Eq.~\eqref{spec_tot} absorption (emission) is described by negative (positive) values of $\tilde{S}(\omega,\tau)$.
In Eq.~\eqref{response} we have introduced the variable time delay $\tau$ between the NIR and the XUV pulses in the incoming field. It should be noted that the ATAS spectrum of Eq. \eqref{response} is not the full intensity spectrum, but the modification of the intensity spectrum as compared to free propagation of the laser pulses. We refer to the full intensity spectrum of ATAS experiments, given by Eq. \eqref{spec_tot}, as the ATAS intensity spectrum.
For the ATAS spectrum to give information about the modification of the XUV pulse, it is used that the frequency profile of the NIR field is well-separated from the frequency profile of the XUV field.
Furthermore, the frequency profile of the higher-order harmonics of the NIR field must also be well separated from the spectral range of the XUV pulse.
We insure that this is the case by choosing NIR pulses with moderate intensities such that only few-photon transitions are induced by these.

In closing this section we relate our expressions for the ATAS spectrum and intensity spectrum to other expressions used in the literature for similar quantities. In Ref.~\cite{gaarde2011transient} (see also Ref. \cite{chen2012light}) the response function
\begin{align}
\tilde{S}(\omega)_\text{Ref.~\cite{gaarde2011transient}}&=4\pi\text{Im} \left[ E_\text{in}^*(\omega) d(\omega) \right] \nonumber \\
&=-\frac{c}{\omega n} \tilde{S}(\omega,\tau)
\label{ref16}
\end{align}
is used as a measure of the modification of the ATAS intensity spectrum. 
In Eq.~\eqref{ref16}, $d(\omega)$ is the Fourier component $d(\omega)=\frac{1}{2\pi}\int_{-\infty}^{\infty} \langle d(t) \rangle e^{i\omega t} dt $.
Apart from the sign convention, our approach to calculate the ATAS spectrum is therefore very similar to the approach of Ref.~\cite{gaarde2011transient}.
Another widely used approach to calculate the ATAS intensity spectrum is to use Beers law (see, e.g., Eq.~(24) in Ref.~\cite{santra2011theory})
\begin{align}
\vert E_\text{XUV} (L,\omega) \vert^2 =& \vert E_\text{XUV} (0,\omega) \vert^2 \nonumber \\ &\exp \left[  -\frac{4\pi \omega}{c} L \text{Im} \left( \frac{P(\omega)}{E_\text{XUV}(\omega)} \right) \right],
\label{Beer}
\end{align}
where $E_\text{XUV}(L,\omega)$ is the Fourier component of the XUV field when it has propagated the distance $L$ into a homogeneous target medium, and $P(\omega)$ is the Fourier component of the polarization. For weak absorption it is accurate to expand Eq.~\eqref{Beer} to first-order in the argument of the exponential function, and we find that
\begin{align}
\vert E_\text{XUV} (L,\omega) \vert^2 \simeq \vert E_\text{XUV} (0,\omega) \vert^2 + L \tilde{S}(\omega,\tau).
\end{align}
This means that both here, and in the theory of Ref.~\cite{santra2011theory}, the ATAS spectrum $\tilde{S}(\omega,\tau)$ of Eq.~\eqref{response} is a good measure of the modification of the XUV intensity spectrum when the generated field is weak ($E_\text{gen} \ll E_\text{in}$).
Some times the cross section rather than the intensity modification is reported in the literature. This is, e.g., the case in Refs. \cite{ott2014reconstruction,PhysRevA.91.013414}. In Ref. \cite{gaarde2011transient} it is shown that the absorption cross section $\sigma (\omega,\tau)$ is directly related to $\tilde{S}(\omega,\tau)$ through $\sigma (\omega,\tau)=-2/(n \vert E_\text{in}(\omega) \vert^2) \tilde{S}(\omega,\tau)$.

\section{\label{results} Results for 1D ${\text{H}_2}^+$ based on TDSE calculations}
In this section we present the ATAS spectrum of Eq.~\eqref{response} for two different 1D models of ${\text{H}_2}^+$.
In Sec.~\ref{fixeds} we present the ATAS spectrum calculated in the fixed nuclei model of ${\text{H}_2}^+$.
For this system we recognize many features from atomic spectra previously reported in the literature
\cite{goulielmakis2010real,ott2014reconstruction,wang2010attosecond,chini2012subcycle,wirth2011synthesized,
pfeiffer2012transmission,wu2013time,chen2012light,chen2013quantum,chini2014resonance,
baggesen2012theory,gaarde2011transient,santra2011theory,PhysRevA.91.013414}. In Sec.~\ref{moving} we present results for the moving nuclei model of ${\text{H}_2}^+$, and we find that dissociation processes completely change the spectrum.

We define the $z$-component of the electric field $E(t)=-\partial_t A(t)$ through the 1D  vector potential
\begin{align}
A(t)=A_0 \exp \left[ -\frac{(t-t_c)^2}{T^2/4} \right] \cos \left[ \omega (t-t_c) \right], \nonumber \\ \quad T=N_c T_c=N_c\frac{2\pi}{\omega} ,
\label{vec_pot}
\end{align}  
where $A_0=E_0/\omega$. $E_0$ relates to the intensity as $I=\vert E_0 \vert^2$, with one atomic unit of intensity equal to $3.51 \times 10^{16}$ W/cm$^2$. In Eq.~\eqref{vec_pot} $t_c$ is the center of the pulse and $\omega$ is the carrier frequency.
The full width at half maximum (FWHM) duration $T_\text{FWHM}$ is related to $T$ used in Eq.~\eqref{vec_pot} by $T_\text{FWHM}=\sqrt{\ln (2)} T$.
For the calculations reported in this section, we use a 700 nm NIR pulse with $T=$ 14.0 fs [$N_c=$ 6 in Eq.~\eqref{vec_pot}] and a 50 nm XUV pulse with $T=$ 330 as [$N_c=$ 2]. The intensities of the pulses are $I_\text{XUV}=5 \times 10^7 $ W/cm$^2$ and $I_\text{NIR}=2 \times 10^{12}$ W/cm$^2$, respectively. 
At these intensities the NIR pulse alone cannot excite the molecule from its ground state to the spectral region of the XUV pulse, as many-photon ($\geq 3$)  processes are greatly suppressed. To propagate solutions of the TDSE for the two systems we use a fast Fourier transform (FFT) split-step operator algorithm \cite{feit1982solution}, as described in Refs.~\cite{Yue13,yue2014dissociative}.

\subsection{\label{fixeds} Fixed nuclei model}
We consider a model for ${\text{H}_2}^+$ of reduced dimensionality with the fixed internuclear distance $R_0=2.07$ which is the expectation value of the internuclear distance in the ground state. 
The TDSE in the dipole approximation and velocity gauge reads 
\begin{equation}
  i\partial_t \Psi(z;R_0,t)=H(t) \Psi(z;R_0,t)
  \label{eq:TDSE_fixed}
\end{equation}
with the Hamiltonian
\begin{equation}
  \begin{aligned}
    H(t)
    =T_\text{e}+V_\text{eN}(z;R_0)+V_\text{N}(z;R_0)+V_\text{I}(t),
    \label{eq:Hamiltonian_fixed}
    \end{aligned}
\end{equation}
where  $z$ is the electronic coordinate  measured with respect to the center-of-mass of the nuclei. The components of the Hamiltonian in Eq.~\eqref{eq:Hamiltonian_fixed} are $T_\text{e}=-(1/2\mu)\partial^2/\partial z^2$, $V_\text{eN}(z,R)=-1/\sqrt{{(z-R/2)^2+a_0}}-1/\sqrt{(z+R/2)^2+a_0}$, $V_\text{N}(R)=1/R$, $V_\text{I}(t)=-i \beta A(t) \partial/\partial z$, $\mu=2m_\text{p}/(2m_\text{p}+1)$ is the reduced electron mass, $\beta = (m_\text{p}+1)/m_\text{p}$, $m_\text{p}=1.836\times 10^3$ is the proton mass, and the softning parameter $a_0=1.508$ for the Coulomb singularity is chosen to produce the exact three-dimensional BO electronic energy at $R_0$.
A box size of $\vert z \vert \leq 100$, $\Delta z=0.39$ and time steps of $\Delta t=5 \times 10^{-3}$ ensured converged results. At the box boundaries a complex absorbing potential was used to avoid unphysical reflections (see Ref.~\cite{Yue13} for details).

\begin{figure}
\includegraphics[width=0.48\textwidth]{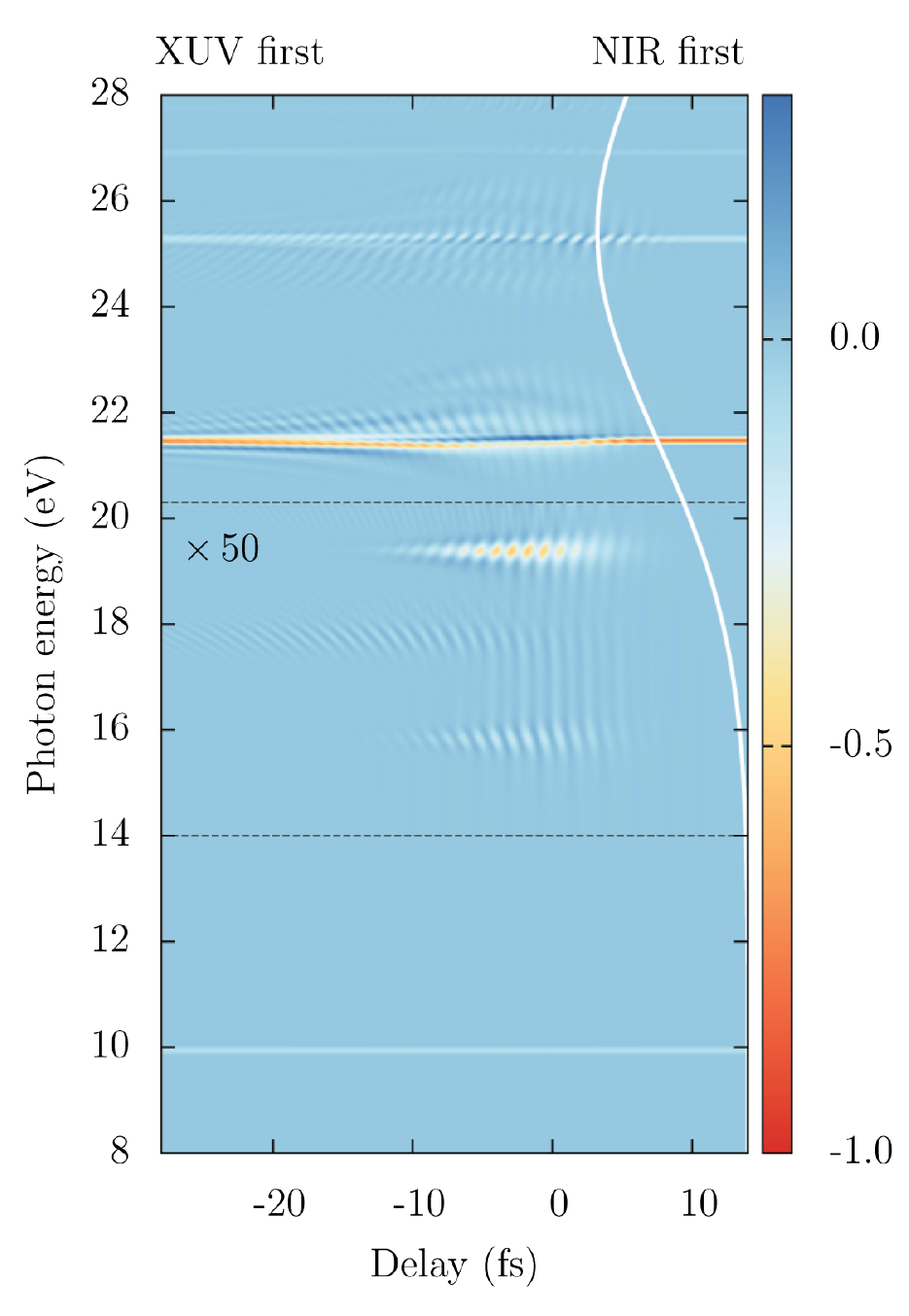}
\caption{\label{Fixed} (Color online) ATAS spectrum $\tilde{S}(\omega,\tau)$ [Eq.~\eqref{response}] calculated for the fixed nuclei model of ${\text{H}_2}^+$, as a function of the photon energy, for a range of delays between the XUV and NIR pulses. The ATAS spectrum is normalized such that maximum absorption is at $\tilde{S}(\omega,\tau)=-1.0$. For photon energies between 14.0 eV and 20.3 eV the ATAS spectrum has been multiplied by a constant factor of 50 to highlight the LIS doublet in that region (see text). The white line in the upper right part of the figure indicates the intensity profile of the XUV pulse. Pulse parameters: $T_\text{XUV}=330$ as, $T_\text{NIR}=14.0$ fs, $\lambda_\text{XUV}=50$ nm, $\lambda_\text{NIR}=700$ nm, $I_\text{XUV}=5\times 10^7$ W/cm$^2$, $I_\text{NIR}=2\times 10^{12}$ W/cm$^2$.}
\end{figure}

Following excitation by the XUV pulse, an oscillating dipole moment is established in the target molecule.
In experiments the time-dependent dipole moment dephases during a time $T_0$ due to collisional broadening and finite detector resolution \cite{chen2013quantum}. To include these effects we use a window function $W(t)$ to damp $\langle d(t) \rangle$ on a time scale, that produces spectra comparable to experimental results. Further, we ensure that the dephasing time is long enough for the qualitative features of the ATAS spectrum not to vary with $T_0$. For a window function
\begin{align}
W(t)=\begin{cases} 1 &(t<\tau) \\ \frac{1}{2} \left\{ 1 + \cos \left[ \frac{\pi ( t-\tau ) }{T_0} \right] \right\} & (\tau \leq t \leq \tau+T_0) \\ 0 &(t>\tau + T_0), \end{cases}
\label{Window}
\end{align}
where $t$ is measured with respect to the center of the NIR pulse, we find that $T_0=3T_{\text{NIR}}\simeq 42$ fs is suitable.
In \fref{Fixed} we present the ATAS spectrum $\tilde{S}(\omega,\tau)$ [Eq.~\eqref{response}] calculated for the fixed nuclei model of ${\text{H}_2}^+$, with delays of the XUV pulse with respect to the NIR pulse varying from -28 fs to 14 fs.
For large positive delays, the NIR pulse arrives at the target well before the XUV pulse. Because the relatively weak NIR pulse leaves the system in its field-free ground state, it will not have any effect on the ATAS spectrum, when it arrives well before the XUV pulse. The spectrum therefore shows clean absorption lines without side-bands or other modifications for delays larger than $\sim 8$ fs. The position of the four absorption lines in \fref{Fixed} at approximately 10 eV,  21.5 eV, 25.5 eV, and 27 eV, correspond to resonances between the field-free ground state and states of opposite symmetry under the reflection $z \to -z$ with energies 9.94 eV,  21.48 eV, 25.29 eV and 26.94 eV above the ground state energy, respectively.

For large negative delays, the XUV pulse arrives well before the NIR pulse. The NIR pulse therefore arrives when an oscillating dipole moment has already been induced. Even though the NIR pulse does not excite additional population from the ground state to the states reached by the XUV field, the NIR pulse will modify $\langle d(t) \rangle$.
One effect of this modification is seen around the absorption peak at $\sim 21.5$ eV in \fref{Fixed}. For delays more negative than $\sim -5$ fs, we see that the absorption line has several sidebands. 
Such sidebands have previously been observed in atomic systems (see, e.g., Ref.~\cite{chen2013quantum} and references therein).
Another effect introduced by the NIR field is a periodic modulation of the absorption line at $\sim 25.5$ eV in \fref{Fixed} (Enlarged in the upper panel of \fref{fringes}). 
This effect extends from large negative delays and all the way to the region where the two pulses overlap in time, and is a consequence of quantum interference between populated excited states of the same symmetry \cite{chen2013quantum,chini2014resonance}. 
For large negative delays the XUV and NIR pulses are separated in time, and the role of the XUV pulse is to populate the excited states at time $\tau$ while the NIR pulse modifies the populations and coherences at a later time t$_0$.
In Appendix~\ref{app} we describe a (three-level) model which explains and reproduces the periodic modulation of the absorption line at $\sim$ 25.5 eV. A result of this treatment is that oscillations of this type always have a period of $\sim T_c^\text{NIR}/2$ with $T_c^\text{NIR}$ the period of a NIR pulse cycle.

The region $-15 \, \text{fs} \lesssim \tau \lesssim 5 \, \text{fs}$, where the two pulses overlap in time, is dominated by intense coupling between states. As a consequence, absorption, and to a smaller extent emission, appears at certain frequencies in this region only. 
Here we will focus on two features both visible in \fref{Fixed}: The Autler-Townes splitting \cite{autler1955stark} of the absorption line at $\sim 21.5$ eV, and the LIS \cite{chen2012light} doublet with spectral features at $\sim 19.4$ eV and $\sim 15.9$ eV corresponding to energies one photon energy above and one photon energy below the even parity state at $17.65$ eV.
When the two pulses overlap in time, we see a clear broadening and a splitting of the absorption line at $\sim 21.5$ eV. 
Such features originate from coupling between bound states and have been observed both experimentally \cite{chini2013sub} and in calculations \citep{chen2013quantum}. As discussed in Refs.~\cite{pfeiffer2012transmission, chen2012light}, the nature of the Autler-Townes splitting can be understood in few-state models.
LIS are signatures of population transfer, by two-photon processes, from the ground state to an excited electronic state of the same symmetry, under the reflection $z \to -z$.
In spectroscopy such a state is often referred to as a dark state. 
Two different processes, one involving absorption of a NIR photon and one involving emission of a NIR photon, can populate the same dark state. As a result LIS always appear in pairs with an energy separation of $2\omega_\text{NIR}$ between the individual absorption features.
The two intermediate states absorbing XUV light are always in two NIR-photon resonance resulting in strong coupling between the states \cite{chen2013quantum}. This coupling leads to a $T_c^\text{NIR}/2$ periodic modulation of the LIS.
In Ref.~\cite{chen2012light} it has been shown that specific LIS in helium can be reproduced using only three states in the calculation of the ATAS spectrum. For ${\text{H}_2}^+$ with fixed nuclei we set up a similar three-level model including the ground state, a dark state $17.65$ eV above the ground state, and a dipole allowed state $21.48$ eV above the ground state. We will refer to these field-free states as $\phi_g (z;R_0)$ (ground state), $\phi_d (z;R_0)$ (dark state), and $\phi_e (z;R_0)$ (dipole allowed, excited state) and their corresponding energies as $E_g=0$, $E_d$, and $E_e$. The wave function of a general state in the three-level model is given by
\begin{align}
\Psi(z;R_0,t) &=c_g(t) \phi_g(z;R_0) + c_d(t) \phi_d(z;R_0) e^{-iE_d (t-\tau)} \nonumber \\ &+ c_e(t) \phi_e(z;R_0)  e^{-iE_e (t-\tau)}.
\label{Two-level}
\end{align}
We find that the three-level model reproduces the LIS shown in \fref{Fixed} at $E_d \pm \omega_\text{NIR}$ well. For the state \eqref{Two-level} the expectation value of the dipole operator is well approximated by
\begin{align}
\langle d(t) \rangle \simeq 2 \text{Re} \left[ d_{ge} c_e(t) e^{-iE_e (t-\tau)} \right],
\label{2-dipole}
\end{align}
since the condition $c_g(t) \simeq 1 >> \vert c_e(t) \vert > \vert c_d(t) \vert$ holds for field parameters used in this work. 

In the pursuit of a simple analytical theory, capable of explaining the characteristics of LIS, we now simplify our three-level model further. This analytical result will be needed for understanding the moving nuclei results of  Sec.~\ref{H2}.
The spectral widths of LIS are determined by the duration of the NIR pulse. This indicates that the temporal region where the XUV field is on only has a small direct effect on the ATAS spectrum (The Fourier transform of the dipole moment in this region is vanishing when compared to the contribution from the remaining temporal region of the NIR field). 
The main contribution from the temporal region where the XUV and NIR pulses are overlapping is therefore to populate the states $\phi_d (z;R_0)$ (two-photon process) and $\phi_e (z;R_0)$ (one-photon process).
When the shorter XUV pulse has died out, the population transfer from $\phi_g (z;R_0)$ to the excited states will stop, and the dynamics will be dominated by coupling between $\phi_d (z;R_0)$ and $\phi_e (z;R_0)$ induced by the NIR field. 
When we consider instants of time after the XUV pulse ($t>\tau$) but before the end of the NIR pulse, the dynamics of the system is approximately governed by the equations
\begin{align}
&\dot{c}_g(t)=0 \label{coef_one} \\
&\dot{c}_d(t)=-iV_{de}(t) e^{i(E_d-E_e) \times (t-\tau)} c_e(t), \label{coef_two} \\
&\dot{c}_e(t)=-iV_{ed}(t) e^{i(E_e-E_d) \times (t-\tau)} c_d(t) \label{coef_three}.
\end{align}
In Eqs.~\eqref{coef_one}-\eqref{coef_three} the multi-photon NIR-coupling between the ground state and excited states has been neglected since such multi-photon processes are greatly suppressed for the field parameters used in this work. 
If we for simplicity assume that the NIR field is given by the monochromatic field $E(t)=E_\text{NIR}^0 \sin (\omega_\text{NIR} t)$, we have
\begin{align}
V_{de}(t)=V_{ed}(t)=-\frac{E_\text{NIR}^0 d_{de}}{2i} \left[ e^{i\omega_\text{NIR}t} - e^{-i\omega_\text{NIR}t} \right].
\end{align}
For the field parameters used in \fref{Fixed} the detuning between $\phi_d (z;R_0)$ and $\phi_e (z;R_0)$ is large ($E_e-E_d=3.83$ eV, $\omega_\text{NIR}=1.77$ eV) and the population transfer is therefore low. 
An approximate analytic expression for $c_e(t)$ can therefore be found using perturbation theory. 
First order perturbation theory can reproduce the characteristics of the LIS and is desirable since the result is relatively simple
\begin{align}
c_e^{(1)}(t) &= -i\int_{\tau}^t dt' V_{de}(t') e^{i(E_e-E_d)\times(t'-\tau)} c_d(\tau) \nonumber \\
&= \frac{E_\text{NIR}^0 d_{de}}{2i} \left[ \frac{e^{i(E_e-E_d+\omega_\text{NIR})\times(t-\tau)}-1}{E_e-E_d+\omega_\text{NIR}}e^{i\omega_{\text{NIR}}\tau} \right. \nonumber \\ &- \left. \frac{e^{i(E_e-E_d-\omega_\text{NIR})\times(t-\tau)}-1}{E_e-E_d-\omega_\text{NIR}}e^{-i\omega_\text{NIR}\tau} \right] c_d (\tau).
\label{e_coef}
\end{align}
If we insert Eq.~\eqref{e_coef} into Eq.~\eqref{2-dipole}, we see that $\langle d(t) \rangle$ will contain terms oscillating with frequencies $E_d \pm \omega_\text{NIR}$. $\tilde{S}(\omega)$ [see Eq.~\eqref{response}] found using this dipole moment, will therefore contain spectral features at $E_d \pm \omega_\text{NIR}$ - the LIS seen in \fref{Fixed}. 
It should be noted that the part of the time-dependent dipole moment responsible for the LIS depends critically on the initial population of the dark state, which is in compliance with the two-photon nature of LIS.
From Eqs.~\eqref{e_coef} and \eqref{2-dipole} we conclude that the NIR-coupling of the excited states in our three-level model gives rise to LIS. 
The simple analytical model for LIS established in this section will be the starting point of an investigation of LIS in molecular systems in Sec.~\ref{H2}.

Figure~\ref{Fixed} and the associated discussion recapitulate many of the physical processes already investigated in atoms. The detailed spectral and temporal information about electron dynamics extracted from \fref{Fixed} is a consequence of the long dipole response of the system, allowing for spectral resolution much better than the Fourier limit of the XUV pulse.
With the phenomena discussed in this section in mind, we are ready to explore the effect of molecular vibrations.

\subsection{\label{moving} Moving nuclei model}
In this section we present the ATAS spectrum of Eq.~\eqref{response} for the moving nuclei model of ${\text{H}_2}^+$ using the same field parameters as in Sec.~\ref{fixeds}. 
We consider a simplified model for ${\text{H}_2}^+$ with reduced dimensionality that includes only the dimension that is aligned with the linearly polarized laser pulse \cite{Kulander96,ver2003time,Weixing01}. Within this model, electronic and nuclear degrees of freedom are treated exactly. The center-of-mass motion of the molecule can be separated, such that the TDSE for the relative motion in the dipole approximation and velocity gauge reads 
\begin{equation}
  i\partial_t\Psi(z,R,t)=H(t) \Psi(z,R,t)
  \label{eq:TDSE_mov}
\end{equation}
with the Hamiltonian
\begin{equation}
  \begin{aligned}
    H(t)
    =T_\text{e}+T_\text{N}+V_\text{eN}(z,R)+V_\text{N}(R)+V_\text{I}(t),
    \label{eq:Hamiltonian_mov}
    \end{aligned}
\end{equation}
where the wave function now depends on the internuclear distance $R$ and the electronic coordinate  $z$  measured with respect to the center-of-mass of the nuclei. The components of the Hamiltonian in Eq.~\eqref{eq:Hamiltonian_mov} were defined after Eq.~\eqref{eq:Hamiltonian_fixed}, except the nuclear kinetic energy $T_N=-(1/m_p)\partial^2/\partial R^2$ and $V_\text{eN}(z,R)=-1/\sqrt{{(z-R/2)^2+a(R)}}-1/\sqrt{(z+R/2)^2+a(R)}$, where the softning parameter $a(R)$ for the Coulomb singularity is chosen to produce the exact three-dimensional $1s\sigma_g$ BO potential energy curve \cite{Madsen12,Yue13}. 
A box of $\vert z \vert \leq 100$, $\Delta z=0.39$ in the electronic coordinate together with a box of nuclear coordinates of $0\leq R \leq 20$, $\Delta R=7.8 \times 10^{-2}$ and time steps of $\Delta t=5 \times 10^{-3}$ ensured converged results.
At the box boundaries complex absorbing potentials were used in both electronic and nuclear coordinates to avoid unphysical reflections (see Refs.~\cite{Yue13,yue2014dissociative} for details).

\begin{figure} 
\includegraphics[width=0.48\textwidth]{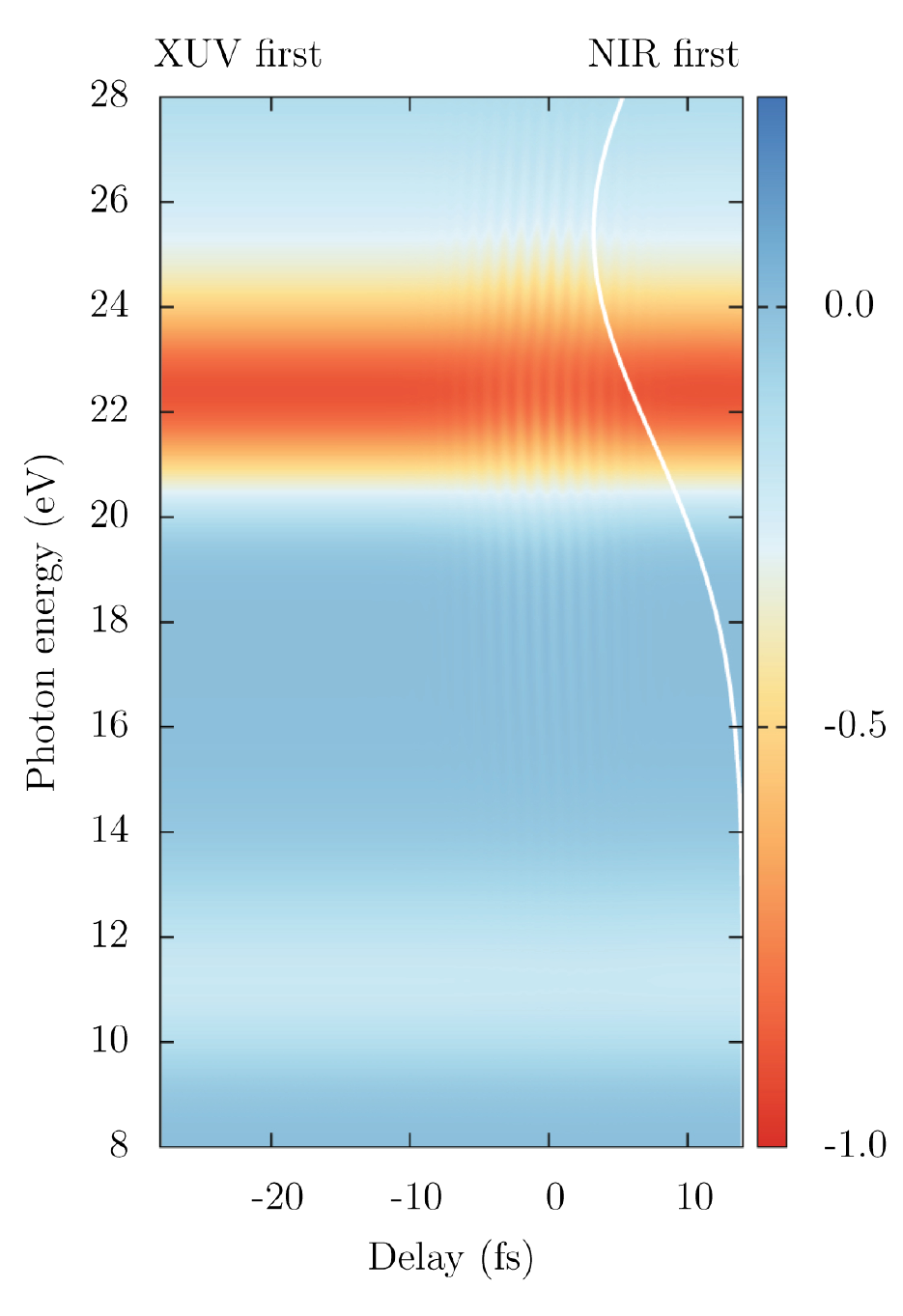}
\caption{\label{Moving} (Color online) Normalized ATAS spectrum $\tilde{S}(\omega,\tau)$ [Eq.~\eqref{response}], calculated using the moving nuclei model [Eqs.\eqref{eq:TDSE_mov} and \eqref{eq:Hamiltonian_mov}], for 1D ${\text{H}_2}^+$, as a function of the photon energy, for a range of delays between the XUV and NIR pulses. The white line to the right shows the intensity profile of the XUV pulse. The pulse parameters are as in \fref{Fixed}.}
\end{figure}

The ATAS spectrum is shown in Fig. \ref{Moving}.
It is clear from \fref{Moving} that the relatively sharp lines we saw for the fixed nuclei model [\fref{Fixed}] have disappeared. 
Two very broad absorption lines corresponding to population transfer between the ground and the first excited state (absorption feature at $\sim 11$ eV) and between the ground state and the third and fifth excited states (merged absorption feature at $\sim 23$ eV) dominate the spectrum.
The broad features in Fig. \ref{Moving} persist as $\tau \to \pm\infty$, and are therefore not a result of the NIR pulse. 
The only effect of the NIR pulse in Fig. \ref{Moving} is the small oscillations with a period of $T_c^\text{NIR}/2$ clearly present between $\sim 14-28$ eV for delays $-5$ fs $\lesssim\tau\lesssim5$ fs, corresponding to the region where the pulses overlap the most. In Sec.~\ref{discussion} we investigate the molecular dynamics responsible for the broad absorption lines in \fref{Moving}. In Sec.~\ref{NS_H2} we also briefly discuss the origin of the $T_c^\text{NIR}/2$ oscillations in \fref{Moving}.

\section{\label{discussion} Analysis and discussion of ${\text{H}_2}^+$ results}
In this section we set up a simplified model to explain the broad absorption features in the ATAS spectrum of the moving-nuclei model of ${\text{H}_2}^+$. To isolate the effect we first neglect the NIR pulse, whose impact we found to be small [\fref{Moving}]. The analysis will show that the dissociative nature of the excited electronic states in ${\text{H}_2}^+$ is responsible for the broad absorption features in the ATAS spectrum and will provide a link between  the spectral width of the ATAS spectrum and the time it takes for the part of the nuclear wave packet found in an excited electronic state to dissociate.
  
\subsection{\label{N-state} $N$-surface model}
In the $N$-surface model the wave function of the molecule is expanded in the lowest $N$ electronic BO states, $\phi_j(z;R)$:
\begin{align}
\Psi (z,R,t) = \sum_{j=1}^N  G_j(R,t) \phi_j(z;R),
\label{state_N}
\end{align}
where $G_j(R,t)$ are the corresponding nuclear wave packets and $z$ denotes the electronic coordinates. In the $N$-surface model of Eq.~\eqref{state_N}, the time-dependent expectation value of the $z$-component of the dipole operator $d$, entering the evaluation of the ATAS spectrum as discussed in Sec.~\ref{theory}, is given by
\begin{align}
\langle d(t) \rangle &= \int dR \int dz \; \Psi^*(z,R,t) d \Psi(z,R,t) \nonumber \\
&=\sum_{i,j}^N \int dR \; G_i^* (R,t) G_j(R,t) d^{\, \text{el}}_{i,j} (R),
\label{dipole_N}
\end{align}
where
\begin{align}
d^{\, \text{el}}_{i,j} (R)=\int dz \; \phi_i^* (z;R) d \phi_j (z;R)
\end{align}
is the electronic dipole moment function. The evolution of the nuclear wave packet $G_i(R,t)$ on the energy-surface $E_{\text{el},i}(R)$, corresponding to the $i$'th electronic state, is governed by the equation
\begin{align} 
i\partial_t G_i(R,t) = \sum_j^N \left( H^{(0)}_i \delta_{i,j} + V_{ij} \right) G_j(R,t),
\label{nuc_evo}
\end{align}
obtained by inserting the $N$-surface expansion~\eqref{state_N} in the TDSE [Eq.~\eqref{eq:TDSE_mov}] and neglecting terms containing derivatives of the electronic states $\phi_j(z,R)$ with respect to $R$. In Eq.~\eqref{nuc_evo} $H^{(0)}_i=-\frac{1}{m_p}\frac{\partial^2}{\partial R^2}+E_{\text{el},i}(R)$ where $E_{\text{el},i}(R)$ is the BO potential energy curve obtained by solving the electronic problem at fixed R. Due to the truncation in the number of electronic states,
the $N$-surface expansion~\eqref{nuc_evo} is not gauge invariant, and the dynamics are only correctly described in the length gauge \cite{giusti1995dynamics}. In Eq.~\eqref{nuc_evo} we therefore have that $V_{ij}=-\beta_{LG} d^\text{el}_{i,j}$, where $\beta_{LG}=(Z_1+Z_2+M_1+M_2)/M_\text{tot}$, with $Z_1$ and $Z_2$ the nuclear charges, $M_1$ and $M_2$ the nuclear masses and $M_\text{tot}$ the total mass of the molecule. Note that an atomic version of the $N$-surface model was used for He in Ref.~\citep{chini2014resonance}. 

For a further analysis of the broad absorption features in \fref{Moving}, we expand the nuclear wave packets $G_i (R,t)$ in the vibrational eigenstates $\chi_{i,k}(R)$ corresponding to the $i$'th electronic state
\begin{align}
G_i(R,t)=\sumint_k dE_{i,k} \; c_{i,k} (t) \chi_{i,k} (R) e^{-i E_{i,k} t},
\label{exp_j}
\end{align}
where $E_{i,k}$ are the eigenenergies of $\chi_{i,k} (R)$. 
The set of $\chi_{i,k} (R)$'s for a specific $i$ can be continuum states, or a mixture of bound and continuum states. The continuum states are energy normalized, such that
\begin{align}
\int dR \chi_{i,p}^*(R) \chi_{i,k}(R) = \delta(E_{i,p}-E_{i,k}).
\end{align}
 Using the expansion~\eqref{exp_j}, the time-dependent dipole moment of Eq.~\eqref{dipole_N} reads
\begin{align}
\langle d(t) \rangle= &\sum_{i,j}^N \sumint_k dE_{i,k} \sumint_p dE_{j,p} \; c^*_{i,k} (t) c_{j,p} (t) \left\langle \chi_{i,k}  \right\vert d_{i,j}^\text{el} \left\vert \chi_{j,p} \right\rangle \nonumber \\ &e^{i(E_{i,k}-E_{j,p})t},
\label{dipole_N2}
\end{align}
where
\begin{align}
\left\langle \chi_{i,k} \right\vert d_{i,j}^\text{el} \left\vert \chi_{j,p} \right\rangle= \int dR \; \chi_{i,k}^* (R) \chi_{j,p} (R) d_{i,j}^\text{el}(R).
\end{align}
Often the function $d^{\, \text{el}}_{i,j} (R)$ varies slowly with $R$, as long as $R$ is varied in a region where $G_i^* (R,t) G_j(R,t)$ is non-zero. In such cases it is a good approximation to let $d^{\, \text{el}}_{i,j} (R)=d^{\, \text{el}}_{i,j} (R_0)$ and we obtain 
\begin{align}
\langle d_-(t) \rangle= &\sum_{i,j}^N \sumint_k dE_{i,k} \sumint_p dE_{j,p} \; c^*_{i,k} (t) c_{j,p} (t) d_{i,j}^\text{el}(R_0) \nonumber \\ &\left\langle \chi_{i,k} \vert \chi_{j,p} \right\rangle e^{i(E_{i,k}-E_{j,p})t},
\label{dipole_N_minus}
\end{align}
where the subscript '-' indicates that the variation in $R$ is neglected for the function $d^{\, \text{el}}_{i,j} (R)$.
In Eq.~\eqref{dipole_N_minus} $\left\langle \chi_{i,k} \vert \chi_{j,p} \right\rangle$ is the Franck-Condon overlap between the states $\chi_{i,k} (R)$ and $\chi_{j,p} (R)$.

The intensity of the XUV pulse is generally very low in ATAS, and it is reasonable to assume that the ground state population remains unchanged during the pulse, i.e., $c_{1,0}(t) \simeq 1$,
and that the expansion coefficients $c_{i,k} (t)$ for vibrational states, corresponding to the BO surfaces of excited electronic states, can be found using first-order perturbation theory
\begin{align}
c_{i,k}^{(1)}(t)= - i\int dt E_\text{in}(t) \left\langle \chi_{i,k} \right\vert d_{i,1}^\text{el} \left\vert \chi_{1,0} \right\rangle  e^{-i\left( E_{1,0} - E_{i,k} \right)t}.
\label{coef_pert}
\end{align}

\subsection{\label{NS_H2}$N$-surface model for ${\text{H}_2}^+$}
In ${\text{H}_2}^+$ all excited BO surfaces are dissociative, and we therefore expect $\langle d(t) \rangle$ to be damped when the excited nuclear wave packets dissociate [see Eq.~\eqref{dipole_N}] This damping leads to a broadening of the ATAS spectrum.
For ${\text{H}_2}^+$ we found that the moving nuclei ATAS spectrum of \fref{Moving} can be reproduced to a very high accuracy using an $N$-surface expansion [Eq.~\eqref{state_N}] of the total wave function followed by propagation of the nuclear wave functions $G_i(R,t)$
in accordance with Eq.~\eqref{nuc_evo}. The nuclear wave packets were propagated using a split step algorithm, $N=6$ in Eq.~\eqref{nuc_evo}, $0\leq R \leq 20$, $\Delta R=7.8 \times 10^{-2}$ and time steps of $\Delta t=5\times 10^{-3}$ ensured converged results. 
The fact that ATAS spectra can be precisely determined in the $N$-surface model is of importance for studies of ATAS in larger systems which can not be treated by direct solution of the TDSE. In Sec.~\ref{H2} we treat H$_2$ in three dimensions using the $N$-surface model.
A very useful property of the $N$-surface expansion is that we can subtract specific surfaces from the expansion during the propagation, and in this way determine their influence on the ATAS spectrum. 
In \fref{Moving} we saw oscillations with a period of $T_c^\text{NIR}/2$ in the region $14-28$ eV. 
By performing separate calculations, we have found that these oscillations are due to two-photon processes involving the electronic even-parity states $\phi_3(z;R)$ and $\phi_5(z;R)$.

In ${\text{H}_2}^+$ the coupling from the ground state to the first excited electronic state is several orders of magnitude larger than the coupling to  higher excited electronic states when evaluated near $R=R_0$. It is therefore expected that pulses with broad frequency distributions primarily populate the first excited state.
\begin{figure} 
\includegraphics[width=0.48\textwidth]{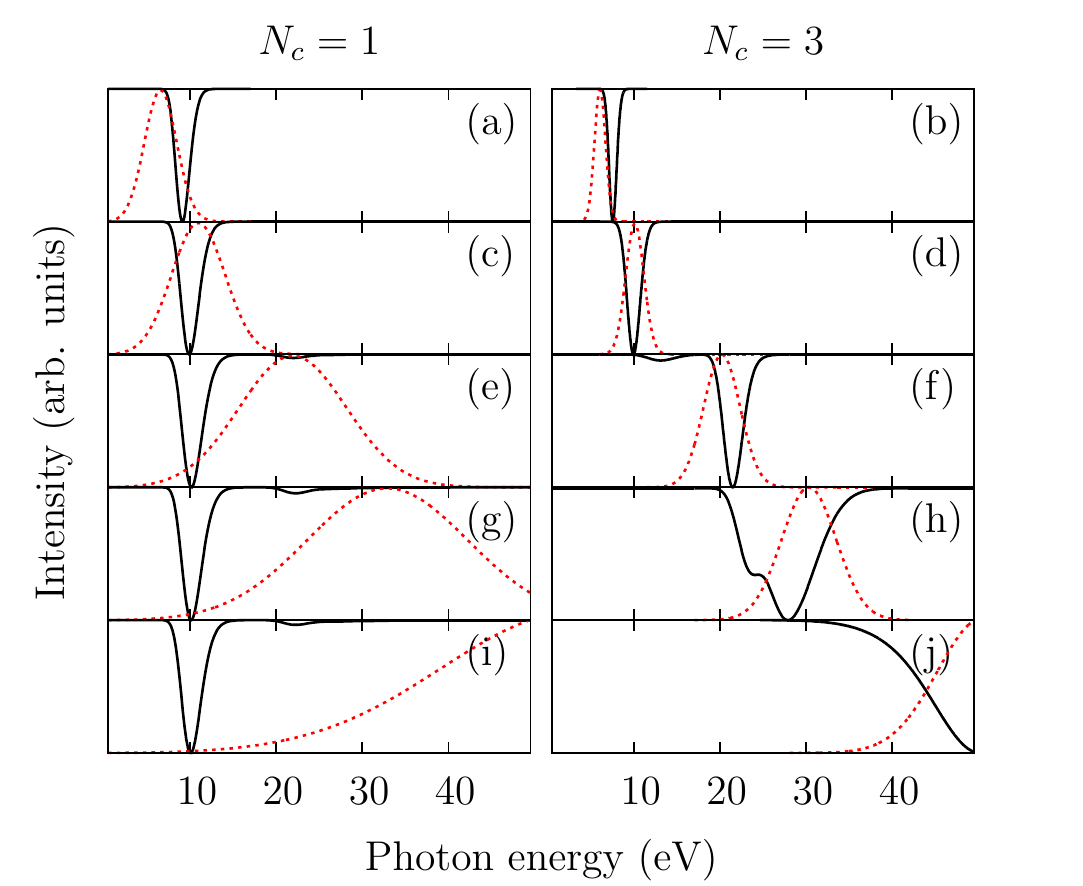}
\caption{\label{Multi} (Color online) ATAS spectrum $\tilde{S}(\omega,\tau)$ [Eq.~\eqref{response}] (solid, black lines) and field intensity (dashed, red lines) as functions of the photon energy. In the left column pulses with $N_c=1$ in Eq.~\eqref{vec_pot} have been used in the calculations, while pulses with $N_c=3$ have been used in the right column. The central carrier frequencies of the pulses used in the calculations are 6 eV for (a) and (b), 10 eV for (c) and (d), 20 eV for (e) and (f), 30 eV for (g) and (h), and 50 eV for (i) and (j).}
\end{figure}
This leads to an interesting effect for ultrashort pulses; the shape of the ATAS spectra shown in \fref{Multi} for pulses with $N_c=1$ in Eq.~\eqref{vec_pot} and carrier frequencies corresponding to energies higher than 20 eV, turn out to be remarkably unchanged as the carrier frequency is varied. As can be seen from \fref{Multi}, this is not the case for longer pulses. The FWHM of the three pulses with $N_c=1$ and central carrier frequencies corresponding to 20 eV, 30 eV and 50 eV are 2.67 eV, 2.67eV and 2.64 eV, respectively.
This indicates that the ATAS spectra for these ultrashort pulses are only determined by the system, and that only the ground state $\phi_1(z;R)$ and the lowest excited electronic state $\phi_2 (z;R)$ are substantially populated. In the case of an ultrashort pulse the dipole signal is therefore expected to be well approximated by a simplified version of Eq.~\eqref{dipole_N_minus}
\begin{align}
\langle d_{-}^{(1)}(t) \rangle=&2 \text{Re} \bigg[ -i \int dE_{2,k} U(E_{2,k}) d_{2,1}^\text{el}(R_0)  \nonumber  \\ & \left\vert \left\langle \chi_{2,k} \vert \chi_{1,0} \right\rangle \right\vert^2 e^{i(E_{1,0}-E_{2,k})t} \bigg], \label{Dipole_pertb2} \\ \nonumber &U(E_{2,k})=\int^t dt' E_\text{in}(t') e^{-i(E_{1,0} -E_{2,k})t'},
\end{align}
where the superscript '$(1)$' indicates that we have used first-order perturbation theory [Eq.~\eqref{coef_pert}] to determine the coefficients in Eq.~\eqref{dipole_N_minus}. 
Equation~\eqref{Dipole_pertb2} is a good approximation for the time-dependent dipole moment as confirmed by the results in \fref{Dipole_mov}, which will be discussed later.

After the pulse is over $U(E_{2,k})$ does not vary with time. In ${\text{H}_2}^+$ the FWHM $\Delta E$ of $\left\vert \left\langle \chi_{2,k} \vert \chi_{1,0} \right\rangle \right\vert^2$ is $\sim 2.5$ eV. For pulses of duration much less than $\frac{2\pi}{\Delta E}=1.7$ fs we can therefore neglect the variation of $U(E_{2,k})$ with respect to $E_{2,k}$; $U(E_{2,k})=U$. 
For such short pulses we have
\begin{align}
\langle d_{-}^{(1)}(t) \rangle &= 2 \text{Re} \bigg[ -i U d_{2,1}^\text{el}(R_0) e^{iE_{1,0}t} \int dE_{2,k}  \nonumber  \\ & \left\vert \left\langle \chi_{2,k} \vert \chi_{1,0} \right\rangle \right\vert^2 e^{-i E_{2,k}t} \bigg], \qquad(t\gg t_c)
\label{Dipole_pertb3}
\end{align}
where $t_c$ is the center of the pulse.
$\int dE_{2,k} \left\vert \left\langle \chi_{2,k} \vert \chi_{1,0} \right\rangle \right\vert^2 e^{-i E_{2,k}t}$ is the inverse Fourier transform of the Frank-Condon overlaps $\left\vert \left\langle \chi_{2,k} \vert \chi_{1,0} \right\rangle \right\vert^2$ with respect to the Fourier variable $E_{2,k}$. It turns out that $\left\vert \left\langle \chi_{2,k} \vert \chi_{1,0} \right\rangle \right\vert^2$ is well approximated by a Gaussian function $D(E_{2,k}-E_2(R_0))$ centered around $E_{2,k}-E_2(R_0)=0$, where $E_2(R_0)$ is the value of the first excited BO surface, evaluated at $R_0=2.07$. Substitution with the variable $\tilde{E}_{2,k}=E_{2,k}-E_2(R_0)$ in Eq.\eqref{Dipole_pertb3} gives
\begin{align}
\langle d_{-}^{(1)}(t) \rangle &= 2 \text{Re} \bigg[ -i U d_{2,1}^\text{el}(R_0) e^{i[E_{1,0}-E_2(R_0)]t} \nonumber \\ &\int d\tilde{E}_{2,k} D(\tilde{E}_{2,k}) e^{-i \tilde{E}_{2,k}t} \bigg].  \qquad(t\gg t_c)
\label{Dipole_pertb4}
\end{align}
Since the Fourier transform of a Gaussian is also a Gaussian, we can now conclude that the time-dependent dipole moment of \eqref{Dipole_pertb4}, after the pulse is over, is oscillating with frequency $E_{1,0}-E_2(R_0)$ and is damped by a Gaussian function. The period of oscillation corresponding to the energy $E_{1,0}-E_2(R_0)$ is $\sim 420$ as.
Figure \ref{Dipole_mov} shows the time-dependent dipole moment from ${\text{H}_2}^+$ calculated using a 30 nm pulse with an intensity of $10^{7}$ W/cm$^2$ and $N_c=1$ in Eq.~\eqref{vec_pot}. The FWHM duration of this pulse is 83 as, and we therefore expect the corresponding dipole signal to be well described by Eq. \eqref{Dipole_pertb4}, when the pulse is over. 
A Gaussian fit to the absolute values of the peaks of the dipole signal, in the region where the pulse has died out (more than 165 as, ie., 2 FWHM durations, after the central time of the XUV attosecond pulse) is shown in \fref{Dipole_mov}, 
and we see that the damping of the dipole signal is well described by a Gaussian function. We also see that the dipole signal is oscillating with a period close to the expected period of 420 as.
\begin{figure} 
\includegraphics[width=0.48\textwidth]{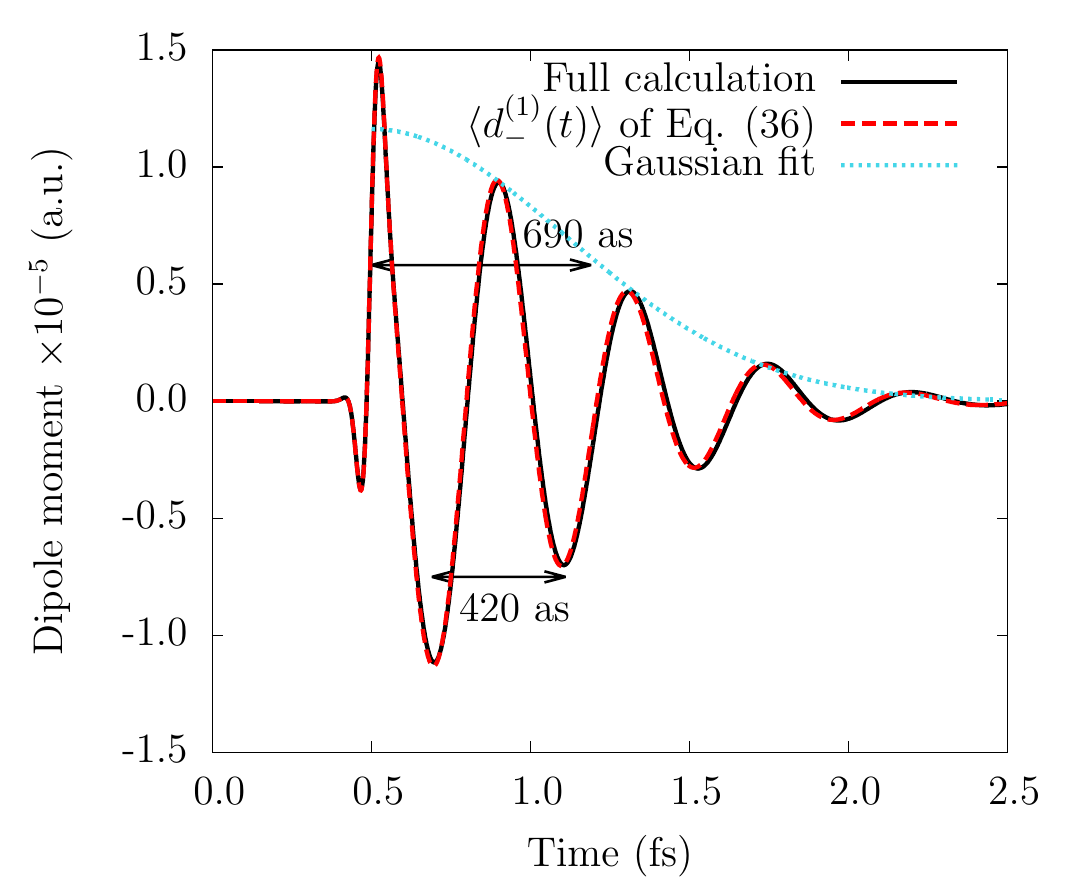}
\caption{\label{Dipole_mov} (Color online) Time-dependent dipole moment from the full TDSE calculation (solid, black line) and the approximation $d_-^{(1)}$ of Eq.~\eqref{Dipole_pertb2} (dashed, red line) calculated for the 1D model of ${\text{H}_2}^+$ and a single-cycle 30-nm pulse. The intensity of the pulse is $10^7$ W/cm$^2$. 
The light blue dotted line shows a Gaussian fit to the norm of the extrema in the damped dipole moment. The first data point included in the fit is the extremum at $t \simeq 0.7$ fs. The XUV pulse is centered at $t=0.5$ fs.}
\end{figure}

Assuming that the Fourier transform of the incoming field is constant in the region where the ATAS spectrum is non-zero (see Fig. \ref{Multi}), the ATAS spectrum of Eq.~\eqref{response} is proportional to the Fourier transform of the time-dependent dipole moment [neglecting the linear frequency dependence from Eq.~\eqref{F_E_gen}].
 For ultrashort pulses the ATAS spectrum of ${\text{H}_2}^+$ is therefore well-approximated by a Gaussian function, and the half-life $t_{1/2}$ of the envelope of $\langle d(t) \rangle$ is related to the FWHM $\omega_\text{FWHM}$ of the ATAS spectrum through the relation
\begin{align}
t_{1/2}=\frac{4 \ln (2)}{\omega_\text{FWHM}}.
\end{align}
From the three ultrashort pulses discussed earlier (left panel in \fref{Multi}) with $\omega_\text{FWHM} \simeq 2.65$ eV we find that $t_{1/2} \simeq 690$ as, which is indicated on \fref{Dipole_mov}. 
The time $t_{1/2}$ indicates the duration it takes the time-dependent dipole-moment to decay and it therefore sets the time scale in which the nuclear wave packet on the excited state $G_2(R,t)$ travels to a region where the overlap with the ground-state wave packet $G_1(R,t)$ is significantly decreased. The value of $t_{1/2}$ shows that this dynamics  takes place on a subfemtosecond time scale. This very fast nuclear motion explains the extremely broad absorption features in the spectrum shown in \fref{Moving}.

\section{\label{H2} H$_2$ in three dimensions}


In ${\text{H}_2}^+$, all BO surfaces corresponding to excited electronic states lead to dissociation. In molecules in general there are also excited BO surfaces supporting bound nuclear motion.
To have a more general description of nuclear motion effects in ATAS, we therefore consider the H$_2$ molecule which is sufficiently simple to allow for an accurate numerical treatment, but still show the characteristics of excited BO surfaces with bound nuclear motion.

Inspired by the good performance of the $N$-surface model in 1D ${\text{H}_2}^+$, we represent the total wave function of 3D H$_2$ in the $N$-surface expansion [Eq.~\eqref{state_N}]. The nuclear wave functions $G_i(R,t)$ are propagated in accordance with Eq.~\eqref{nuc_evo} using a FFT split-step algorithm.
BO surfaces and dipole transition moments used in the propagation are given in the literature \cite{fantz2006franck}.
We must ensure that the basis used in the expansion of Eq.~\eqref{state_N} is large  enough to contain all physics relevant to the ATAS spectrum. 
To reduce the number of basis states $N$ needed, we choose the intensity of the dressing field low enough that processes involving more than two low-energy photons are suppressed.
 Calculations of $\tilde{S}(\omega,\tau)$ for H$_2^+$ with fixed nuclei did not show any signs of 3-NIR-photon or 4-NIR-photon processes for the field parameters we use in this section. To further reduce $N$, we use a 1600 nm IR dressing pulse. Under these criteria it is sufficient to use $N=10$ BO surfaces in the expansion.
The propagation of $G_i(R,t)$ with $0 \leq R \leq 20$, $\Delta R = 7.8 \times 10^{-2}$ and time steps of $\Delta t=5\times 10^{-3}$ ensured convergenced results.  
To highlight the effect of nuclear dynamics on ATAS spectra we also preform a fixed nuclei calculation on 3D H$_2$ using the same number of electronic states as described above. In the fixed nuclei calculation we use an internuclear separation of $R=1.393$, corresponding to the separation at minimum energy in the ground state BO curve.

\begin{figure} 
\includegraphics[width=0.48\textwidth]{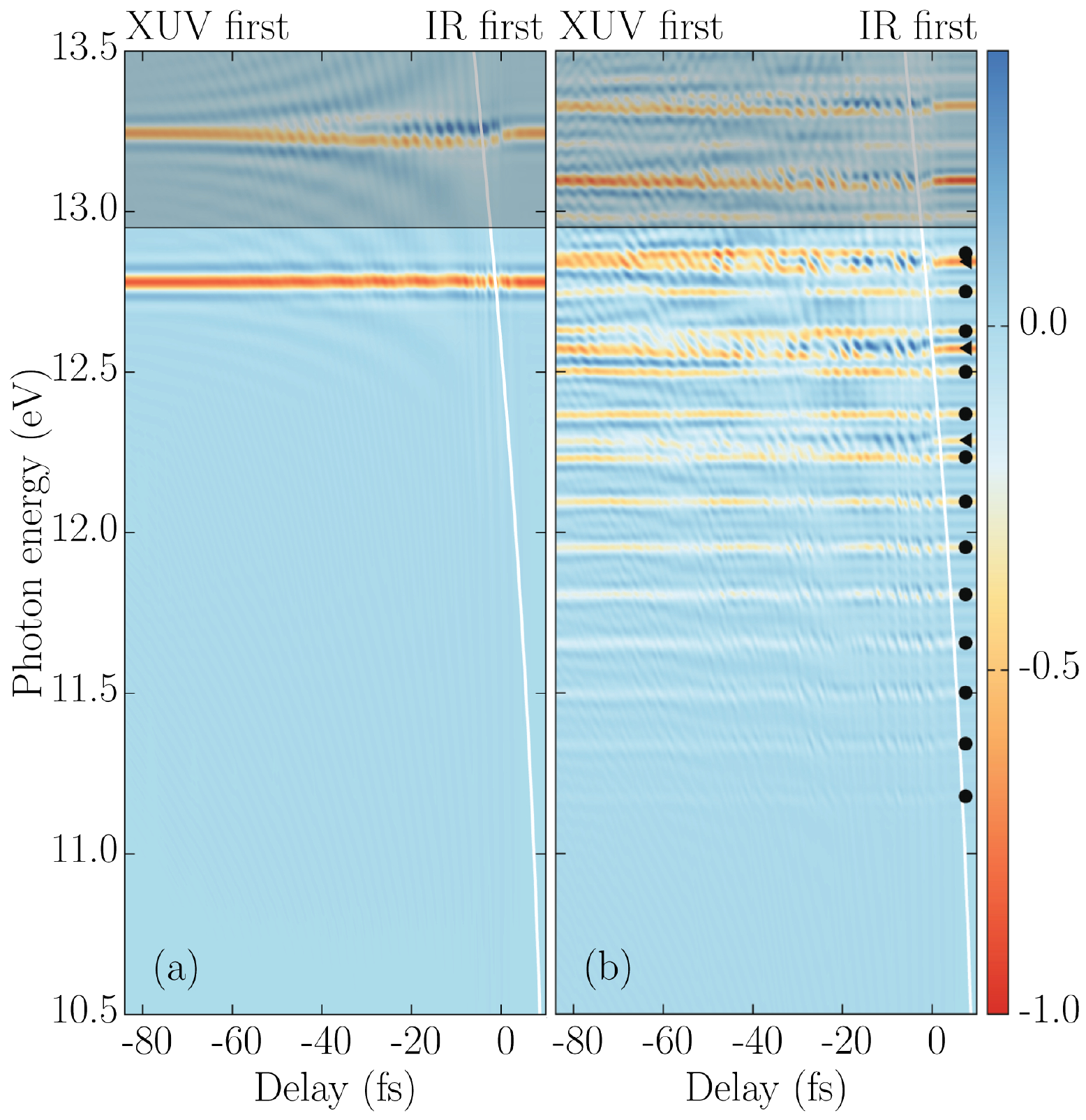}
\caption{\label{1600} (Color online) Normalized ATAS spectrum $\tilde{S}(\omega,\tau)$ [Eq.~\eqref{response}] calculated for 3D H$_2$. (a) for fixed nuclei, (b) for moving nuclei. The shaded areas indicate photon energy regions where the spectra are expected to be influenced by the finite electronic basis (see text). The white line to the right indicates the intensity profile of the XUV pulse. The dots and triangles to the very right in (b) are located at the field-free energies of the vibrational states corresponding to the BO curves B$^u$ (black dots) and C$^u$ (black triangles).
Pulse parameters: $T_\text{XUV}=530$ as, $T_\text{IR}=10.7$ fs, $\lambda_\text{XUV}=80$ nm, $\lambda_\text{IR}=1600$ nm, $I_\text{XUV}=5\times 10^7$ W/cm$^2$, $I_\text{IR}= 10^{12}$ W/cm$^2$.}
\end{figure}

In \fref{1600} we present the ATAS spectrum $\tilde{S}(\omega,\tau)$ calculated for $\text{H}_2$ with delays of the XUV pulse with respect to the IR pulse varying from -84 fs to 10 fs. Figure~\fref{1600}(a) shows the fixed nuclei results, while \fref{1600}(b) shows the ATAS spectrum, when the nuclei are allowed to move. 
The field parameters are the same in the two calculations. The IR pulse is a 1600 nm pulse with a FWHM of $8.83$ fs [$N_c=2$ in Eq.~\eqref{vec_pot}], and the XUV pulse is an 80 nm pulse with a FWHM of $440$ as ($N_c=2$). The intensities of the pulses are $I_\text{IR}=10^{12} $ W/cm$^2$ and $I_\text{XUV}=5 \times 10^{7}$ W/cm$^2$, respectively.
The 11th (singlet) BO surface in H$_2$ has an energy minimum of $\simeq 14.5$ eV above the ground state energy \cite{sharp1970potential}, and the photon energy of the IR field is 0.775 eV. In Fig. 5 regions of photon energies above $14.5$ eV $-2 \times 0.775$ eV $=12.95$ eV are shaded indicating that we expect structures in this region of the ATAS spectrum not to be well described in the basis of 10 BO surfaces. 
In $\text{H}_2$ we, in addition to electronic dynamics, want to follow nuclear dynamics occurring on a much longer timescale. We therefore use a longer dephasing time for the time-dependent dipole moment than for the case of ${\text{H}_2}^+$ with fixed nuclei. For the calculations in this section we use the window function~\eqref{Window} with $T_0=48$ fs, but shift the window function such that $\langle d(t) \rangle$ is undamped the first 73 fs after the end of the XUV pulse.
 
From \fref{1600} we immediately see that the ATAS spectrum of H$_2$ is highly affected by the nuclear dynamics of the system. The effect of nuclear motion on the ATAS spectrum in H$_2$ is, however, very different from what we saw in ${\text{H}_2}^+$. In the fixed nuclei ATAS spectrum of \fref{1600}(a) we only see a single absorption line in the un-shaded region. 
For H$_2$ with a fixed internuclear separation of $R_0=1.393$, the excited states considered have energies higher than the lowest vibrational state in the corresponding BO curve for the moving nuclei system. 
It is therefore expected, that we can go to higher energies in the ATAS spectrum without expanding the basis of the calculation. 
For energies higher than 12.95 eV additional absorption lines are visible in the fixed nuclei ATAS spectrum [one is included in the shaded area of \fref{1600}(b)]. These absorption lines and the absorption line at $\simeq 12.8$ eV all behave very much like the absorption lines in the fixed-nuclei spectrum of ${\text{H}_2}^+$; we observe interference of the type discussed in Appendix A between states in two NIR photon resonance, Autler- Townes splittings of absorption lines etc. 

\begin{figure} 
\includegraphics[width=0.48\textwidth]{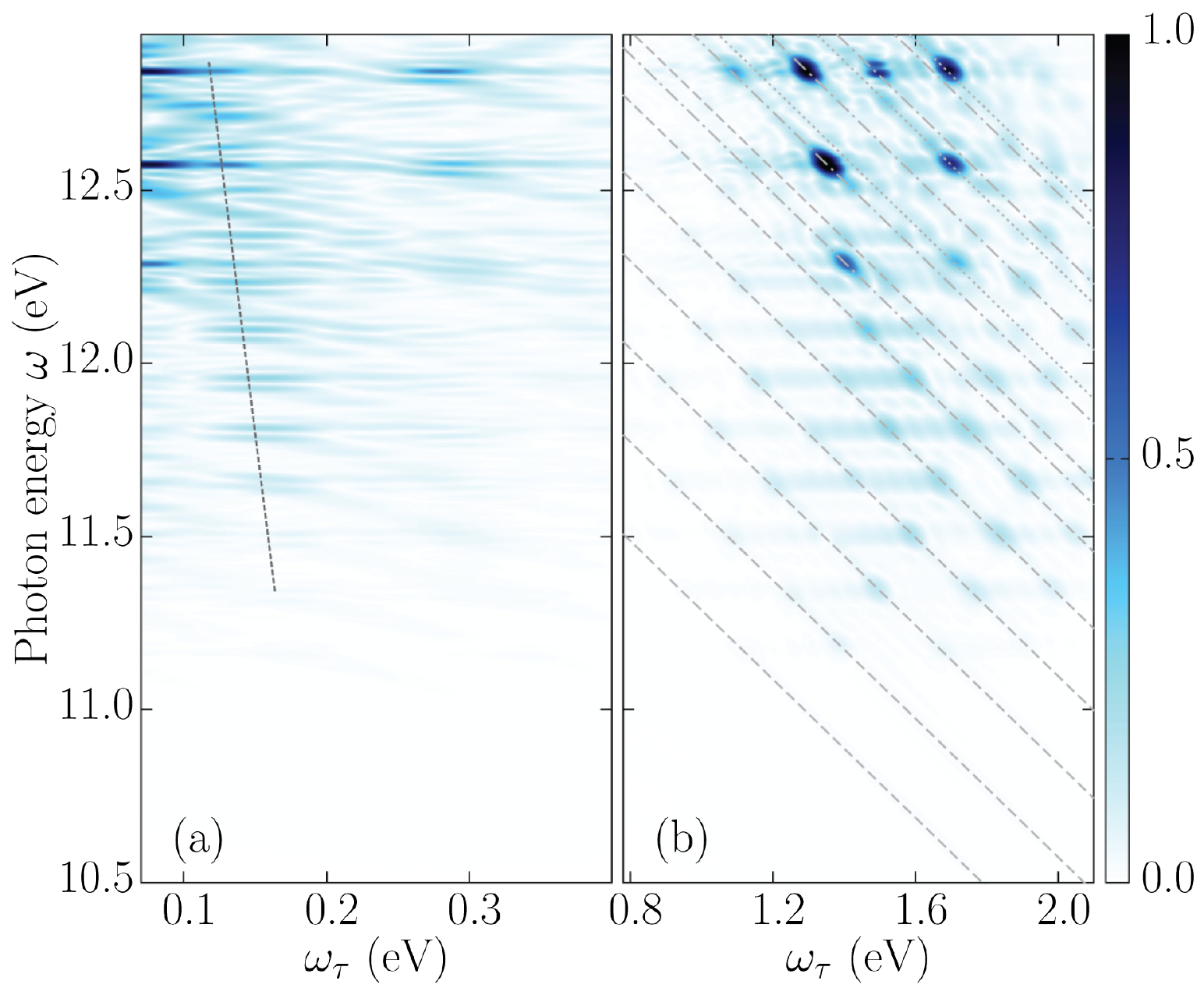}
\caption{\label{F_Freq} (Color online) Absolute value of the Fourier transform~\eqref{F_tau}. (a) for relatively low frequencies and (b) for higher frequencies. The dashed line in (a) indicates the local energy separation between vibrational states in the B$^u$ curve corresponding to the local vibrational period of Eq.~\eqref{T_loc}. 
The lines in (b) are $E_i-\omega_{\tau}$ [see Eqs. (40) and (42)] when $E_i$ is either the energy of one of the seven lowest vibrational states of the C$^u$ surface (dashed lines), one of the five lowest vibrational states of the B'$^u$ surface (dash-dotted lines), or one of the three lowest vibrational states of the D$^u$ surface (dotted lines). See Ref. \cite{fantz2006franck} for information on the energy levels.}
\end{figure}

The moving nuclei ATAS spectrum of H$_2$ shown in \fref{1600}(b) is, on the other hand, very different from both the fixed nuclei ATAS spectrum of H$_2$, and from the moving nuclei ATAS spectrum of ${\text{H}_2}^+$ (\fref{Moving}). Even though only three singlet BO curves (B$^u$, C$^u$ and EF$^g$) are within the energy range considered in \fref{1600}(b), we see rich structures in the ATAS spectrum. 
For the positive delays to the outermost right in \fref{1600}(b) we see that these structures converge towards absorption lines corresponding to population of the field-free vibrational states of the B$^u$ and C$^u$ surfaces.
When the delay between the XUV pulse and the NIR pulse is varied, we see that oscillations occur in the spectrum on two different time scales. Throughout the spectrum there are oscillations on the time scale of half an IR period ($\sim 2.7$ fs), and in the absorption lines corresponding to vibrational states of the B$^u$ surface [indicated by black dots in \fref{1600}(b)], we see additional oscillations on a much longer time scale.
Introducing the Fourier transform of $\tilde{S}(\omega,\tau)$ [Eq.~\eqref{response}]
\begin{align}
\mathcal{F} [\tilde{S}(\omega,\tau)] =\frac{1}{2\pi}\int_{-\infty}^{\infty} \tilde{S}(\omega,\tau) \exp(i\omega_\tau \tau) d\tau,
\label{F_tau}
\end{align}
we are able to analyze the oscillations of \fref{1600}(b) closer. The Fourier transform Eq.~\eqref{F_tau} of the delay-dependent ATAS spectrum is shown in \fref{F_Freq}(a) for low and in \fref{F_Freq}(b) for higher frequencies $\omega_\tau$. 
The low frequency structure in the Fourier transform of the ATAS spectrum is a consequence of nuclear dynamics in the molecule. In the $N$-surface expansion~\eqref{state_N}, we can think of the XUV pulse as generating nuclear wave packets on BO surfaces corresponding to different excited electronic states. If the BO surfaces were harmonic, the wave packets would oscillate periodically with a period of $2\pi/{\omega_h}$, where $\omega_h$ is the frequency of the harmonic oscillator. 
However, the BO curves of $\text{H}_2$ are not perfectly harmonic, and we must account for their anharmonicity.
We do this by defining a local vibrational period for the $n$'th vibrational state as
\begin{align}
T^\text{loc}_{n}=\frac{1}{2} \left( \frac{2\pi}{\Delta\omega_{n-1,n}} +\frac{2\pi}{\Delta\omega_{n,n+1}} \right) \qquad (n>1),
\label{T_loc}
\end{align}
where $\Delta \omega_{k,l}$ is the energy difference between the $k$'th and the $l$'th vibrational states of the same electronic state. 
The local energy separation $2\pi/T^\text{loc}_{n}$ for the $n$'th vibrational state is plotted in \fref{F_Freq}(a). In \fref{F_Freq}(a) we can also vaguely see frequency components corresponding to the energy separation of the vibrational states in the C$^u$ curve around 0.3 eV.

In \fref{F_Freq}(b) we present the higher frequency components of the Fourier transform of the delay-dependent ATAS spectrum. Note that the structures of \fref{F_Freq}(b) are centered around the energy $1.55$ eV of two 1600nm IR photons.
Much of the structure in \fref{F_Freq}(b) can be explained as interference between different paths to the same final state, just as the interference in the absorption lines for fixed-nuclei ${\text{H}_2}^+$ (see Appendix~\ref{app}). 
For both fixed and moving nuclei the oscillation has a period of approximately half a NIR or IR field cycle.
We now focus on one absorption peak in \fref{1600} and denote its energy by $\omega$. 
If the field-free state corresponding to this absorption peak couples to a state with energy $E_i$, the absorption peak at energy $\omega$ will contain a term proportional to $\cos[(\omega-E_i)\tau]$, neglecting a phase. 
We now take the Fourier transformation defined in Eq. \eqref{F_tau}, but with $\tilde{S}(\omega,\tau) \to \cos[(\omega-E_i)\tau]$, and see that we only obtain a nonvanishing contribution when
\begin{align}
E_i \pm \omega_\tau = \omega.
\label{ress}
\end{align}
In \fref{F_Freq}(b) we plot $E_i - \omega_\tau$ for a number of energies $E_i$ corresponding to vibrational states in different BO surfaces. In practice the signal $\cos[(\omega-E_i)\tau]$ does not exist for all $\tau$, and the peaks in \fref{F_Freq}(b) are therefore not delta functions, but rather peaks smeared out around sets of $\omega$'s and $\omega_{\tau}$'s obeying the condition \eqref{ress}. 
If we focus on the absorption feature at $\sim 11.5$ eV we see from \fref{F_Freq}(b) that the corresponding field-free state couples to at least three other vibrational states through the IR field. 
From this analysis we conclude that the interference mechanism resulting in $T_c^\text{IR}/2$ oscillations in the molecular absorption lines of \fref{1600} is very similar to the mechanism resulting in $T_c^\text{NIR}/2$ oscillations in the fixed-nuclei absorption lines discussed in Appendix~\ref{app}. Other processes such as LIS and Autler-Townes splittings of the absorption lines have not been found in the analysis of moving nuclei ATAS spectra of H$_2$. 

\begin{figure} 
\includegraphics[width=0.48\textwidth]{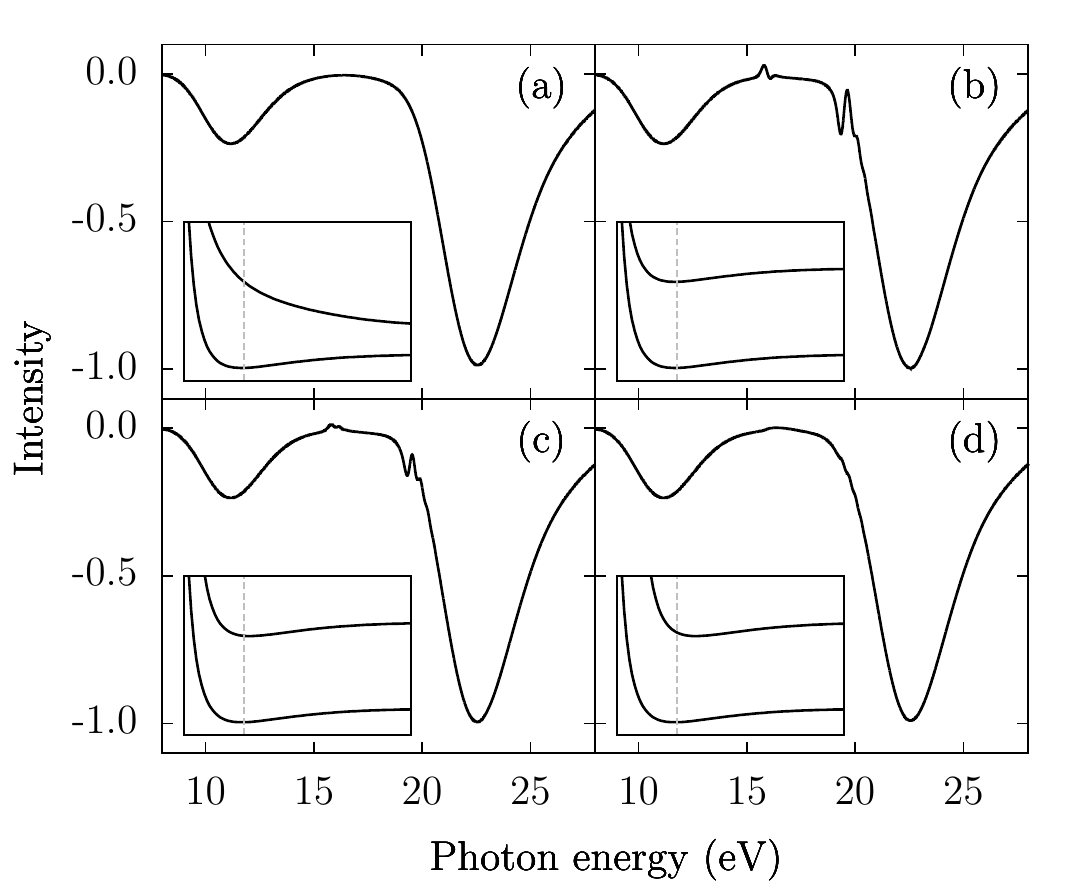}
\caption{\label{LIS} Normalized ATAS spectrum $\tilde{S}(\omega,\tau)$ [Eq.~\eqref{response}] calculated at $\tau=0$ for (a) ${\text{H}_2}^+$ and the artificial ${\text{H}_2}^+$ molecule (see text) with (b) $R_T=0$, (c) $R_T=0.234$ and (d) $R_T=0.625$. The four insets show the BO curves $E_g(R)$ and $E_d(R)$. In the insets $R_0$ is indicated by a dashed gray line. The pulse parameters are as in \fref{Fixed}, but with a fixed delay $\tau=0$ between NIR and XUV pulses.}
\end{figure}

We now investigate whether LIS at well-defined energies are hidden in the rich molecular structure of ATAS spectra in H$_2$, or a more fundamental nuclear motion effect suppresses such LIS in molecular systems.
For this purpose we use a perturbative electronic three-level model similar to the one used in Eqs.~\eqref{Two-level}-\eqref{e_coef} for ${\text{H}_2}^+$ with fixed nuclei, but now accounting for nuclear motion.
The model contains the electronic states $\phi_g (z;R)$ (ground state), $\phi_d (z;R)$ (dark state) and $\phi_e (z;R)$ (essential, dipole allowed, excited state). The wave function of a general state in the model can be written as
\begin{align}
\Psi(z,R,t)=& G_g(R,t) \phi_g (z;R) + G_d(R,t) \phi_d (z;R) \nonumber \\ +& G_e(R,t) \phi_e(z;R) \nonumber \\
=&\chi_{g,0}(R)\phi_g (z;R) \nonumber \\ +& \sumint_l dE_{d,l} c_{d,l}(t) \chi_{d,l} (R) e^{-iE_{d,l}(t-\tau)} \phi_d (z;R) \nonumber \\ +& \sumint_k dE_{e,k} c_{e,k}(t) \chi_{e,k}(R) e^{-iE_{e,k}(t-\tau)} \phi_e (z;R), \label{Nuc_3_state}
\end{align} 
where the nuclear wave packets $G_g(R,t)$, $G_d(R,t)$ and $G_e(R,t)$ have been expanded in their vibrational eigenstates [see Eq.~\eqref{exp_j}]. In Eq.~\eqref{Nuc_3_state} the subscripts $g,d$ and $e$ label the three electronic states, while the subscripts $k$ and $l$ label the vibrational states corresponding to the electronic states $e$ and $d$, respectively. 
In obtaining Eq.~\eqref{Nuc_3_state} we have approximated $G_g(R,t)$ by its vibrational ground state wave function $\chi_{g,0}(R)$ and use $E_{g,0}=0$. The approximation $G_g(R,t) \simeq \chi_{g,0}(R)$ is reasonable for the field parameters used in ATAS and is similar to the approximation $c_g(t) \simeq 1$ made in Sec.~\ref{fixeds}.
Under the approximation $d_{g,e}^{\text{el}}(R)=d_{g,e}^{\text{el}}(R_0)$ [see Eq.~\eqref{dipole_N_minus}] the expectation value of the dipole moment found using Eq.~\eqref{Nuc_3_state} is well approximated by [see Eq.~\eqref{2-dipole}]
\begin{align}
\langle d_-(t) \rangle \simeq & 2 \text{Re} \bigg[ d_{g,e}^\text{el}(R_0) \sumint_k dE_{e,k} \nonumber \\ &\left\langle \chi_{g,0}  \vert \chi_{e,k} \right\rangle \; c_{e,k} (t) e^{-iE_{e,k}(t-\tau)} \bigg]. \label{dipole_three_nuc}
\end{align}
As in Sec.~\ref{fixeds} we approximate the NIR (or IR) field by a monochromatic field 
$E(t)=E^0 \sin (\omega_c t)$ and evaluate the coefficients $c_{e,k} (t)$ by first order perturbation theory
\begin{align}
c_{e,k}^{(1)}(t) =& \frac{E^0  d_{d,e}^\text{el} (R_0)}{2i} \sumint_l dE_{d,l} \nonumber \\ &\bigg[ \frac{e^{i(E_{e,k}-E_{d,l}+\omega_c)\times(t-\tau)}-1}{E_{e,k}-E_{d,l}+\omega_c}e^{i\omega_c\tau} -  \nonumber \\ & \frac{e^{i(E_{e,k}-E_{d,l}-\omega_c)\times(t-\tau)}-1}{E_{e,k}-E_{d,l}-\omega_c}e^{-i\omega_c\tau} \bigg] \langle \chi_{e,k} \vert \chi_{d,l} \rangle c_{d,l}(\tau).
\label{dipole_three_nuc_coef}
\end{align}
Inserting Eq.~\eqref{dipole_three_nuc_coef} into Eq.~\eqref{dipole_three_nuc} shows that the energies $E_{e,k}$ disappear in the final expression for the time-dependent dipole moment, leaving terms oscillating with frequencies $E_{d,l} \pm \omega_c$. 
The distribution and population of vibrational states corresponding to the electronic state $\phi_e (z,R)$ does therefore not affect the dephasing of the dipole moment responsible for LIS.
The distribution and population of vibrational states corresponding to the electronic state $\phi_d (z,R)$ are, on the other hand, crucial for the existence of LIS as spectral structures at well-defined energies. The more dark vibrational states populated, the faster the resulting dipole moment will dephase.
Even in the situation where a finite number of discrete vibrational states are populated, and one would expect a partial revival of the time-dependent dipole moment, the revival is often absent due to the finite duration of the NIR pulse. 
We therefore expect LIS with well-defined energies to be suppressed in the molecular ATAS spectrum when more than a few vibrational dark states are populated. 
To test this hypothesis we return to ${\text{H}_2}^+$ for a short notice.
In ${\text{H}_2}^+$ the BO curve corresponding to the dark electronic state investigated in Sec.~\ref{results} is dissociative, and a continuum of vibrational states are populated. We therefore expect the LIS corresponding to this dark electronic state to be suppressed in the ATAS spectrum.
The ATAS spectrum $\tilde{S}(\omega,\tau)$ using the pulse parameters from \fref{Fixed} and a fixed delay of $\tau=0$ is shown in \fref{LIS}(a).
We now construct an artificial ${\text{H}_2}^+$ molecule by replacing the BO curve $E_d(R)$ corresponding to the dark state with $E_g(R-R_T)-\left[ E_g(R_0) - E_d(R_0) \right]$, for $R_T=0$ this BO curve is parallel to the ground state curve $E_g(R)$.
According to the Franck–-Condon principle only the dark vibrational ground state will be populated when the two BO curves are parallel. 
We therefore expect LIS in the ATAS spectrum of the artificial ${\text{H}_2}^+$ molecule with $R_T=0$, shown in \fref{LIS}(b). 
We see that this is indeed the case since spectral features of the expected width, an energy separation of two NIR photon energies, and centered around the energy of the dark vibrational ground state, appear in the spectrum. 
It should be noted that the coupling to the dark state curve has not been changed, and the results should therefore only be interpreted qualitatively. 
In Figs.~\ref{LIS}(c) and \ref{LIS}(d) the ATAS spectrum is shown for the artificial ${\text{H}_2}^+$ molecule with non-zero $R_T$. 
As $R_T$ is increased, the LIS are damped. 
The reason for this damping is that the slope of the BO curve, corresponding to the dark state, evaluated at $R_0$, increases with $R_T$ and more vibrational dark states are therefore populated for larger $R_T$.
In H$_2$ the dark state curves of interest are far from parallel to the ground state curve, and hence we do not expect any significant LIS in the ATAS spectrum.
In ATAS experiments using a molecular gas target, we therefore predict that LIS will be very difficult to detect. An exception to this general rule might be $\text{N}_2$ where the BO surfaces are remarkably parallel. It should be noted that Eq.~\eqref{dipole_three_nuc} is valid only when the detuning is large, and first-order perturbation theory can be applied. The conclusions, however, remain the same if perturbation theory of a higher order is used. Further, our calculations show that there is no significant difference in the nature of LIS in systems with large and small detunings.

\section{\label{conclusion} Conclusion}
The effect of nuclear motion in ATAS spectra has been elucidated. We performed  calculations in 1D ${\text{H}_2}^+$ and 3D H$_2$ with fixed nuclei, and found that the spectra of these systems are very similar to atomic spectra.
Moreover we investigated the same two systems including nuclear motion. 
For the systems with moving nuclei, we saw that the nature of the BO surfaces, corresponding to excited electronic states, were crucial. 
In ${\text{H}_2}^+$ all excited BO surfaces lead to dissociation, and as a result the ATAS spectrum shows very broad absorption features reflecting the damping of the time-dependent dipole moment by the dissociating nuclei. 
In the theoretical analysis of these broad absorption lines we found that, for specific field parameters, the width of the absorption lines were solely system dependent. 
The width of the lowest absorption line in the ATAS spectrum could therefore be linked directly to the time it takes for a nuclear wave packet to dissociate on the excited dissociative BO curve. Temporal information about this process can be obtained using only one pulse, i.e., without dressing of the system.
In H$_2$ the broad absorption features of ${\text{H}_2}^+$ were replaced by rich structures in the ATAS spectrum. 
For H$_2$ we focused on explaining the oscillations in the absorption lines, and found that these could be put into two different categories; one category was oscillations on the timescale of half the IR period, the other category was oscillations on a much longer timescale. 
We argued that the oscillations on the timescale of half the IR period were very similar in nature to the oscillations seen in the absorption lines of the atomic-like ${\text{H}_2}^+$ with fixed nuclei. We therefore conclude that these oscillations are due to interference between populated states of the same symmetry. The oscillations on longer timescales, however, are a purely molecular signature associated with transfer of population between vibrational states. 
Finally we found that the existence of LIS at well-defined energies in molecular ATAS spectra depends critically on the number of populated vibrational dark states. 
As a result clear, unambiguous detection of LIS can only be expected in systems where the BO curves of the ground state and the involved dark state(s) are nearly parallel, or if one in another way can populate only a few vibrational dark states.

\section*{Acknowledgments}
This work was supported by the ERC-StG (project no. 277767-TDMET), and the VKR center of excellence QUSCOPE.

\appendix
\section{\label{app} Interference between excited states in fixed nuclei ${\text{H}_2}^+$}
In this appendix we model the interference process between excited states of the same symmetry (under $z \to -z$) resulting in a $T_c^\text{NIR}/2$ periodic modulation of the corresponding absorption lines in the ATAS spectrum.
We focus on the interference fringes in the absorption line at $\simeq 25.5$ eV in \fref{Fixed} corresponding to an energy difference $E_6$ between the ground state and the fifth excited state. 
At large negative delays the population of the fifth excited state $\phi_6 (z;R_0)$ has two contributions for $t > t_0$. A direct contribution from the XUV pulse alone, and an indirect contribution from the XUV and NIR pulses acting in a combined manner. In the indirect process the XUV pulse populates the third excited state $\phi_4 (z;R_0)$ at time $\tau$, and at a later time $t_0$ the NIR pulse excites population to the fifth excited state through the absorption of two photons. Keeping only terms that contribute to a dipole signal with frequency $E_6$ we can write up the wave function of the system as
\begin{align}
\Psi(z;R_0,t) =&\phi_1(z;R_0) + \left( c_\text{d} e^{-iE_6(t-\tau)} \right. \nonumber \\ +& \left. c_\text{id} e^{-iE_6(t-t_0)}e^{-iE_4(t_0-\tau)}  \right)\phi_6(z;R_0) \quad (t>t_0)\nonumber \\
=&\phi_1(z;R_0) + \left( c_\text{d}+c_\text{id}e^{i\varphi}\right) \nonumber \\ &e^{-iE_6(t-\tau)} \phi_6(z;R_0) \quad (t>t_0),
\label{simple_psi}
\end{align}
where $\phi_1 (z;R_0)$ is the ground state with energy $E_1=0$, $E_i,i=4,6$ are the energies of $\phi_4 (z;R_0)$ and $\phi_6 (z;R_0)$, $c_\text{d}$ and $c_\text{id}$ are the amplitudes corresponding to the direct and the indirect absorption processes, and the phase $\varphi$ is given by
\begin{align}
\varphi=(E_6-E_4)\times(t_0-\tau).
\label{simple_phi}
\end{align}
This phase is similar to the phase found in Ref.~\cite{chen2013quantum} where constructive interference conditions are investigated.
To obtain Eq.~\eqref{simple_psi} we used that the population of the ground state is almost constant in time due to the low intensity of the XUV pulse. 
The time-dependent dipole moment in the simplified model is given by
\begin{align}
&\langle d(t) \rangle=\langle \psi \vert d(t) \vert \psi \rangle \propto \nonumber \\ &\begin{cases} 0 &(t<\tau) \\ \cos \left[ E_6 (t-\tau) \right] &(\tau<t<t_0) \\ c_\text{d}\cos \left[ E_6 (t-\tau) \right] + c_\text{id} \cos \left[ E_6 (t-\tau )-\varphi \right], &(t>t_0) \end{cases}
\label{simple_dipole}
\end{align}
\begin{figure} [b]
\includegraphics[width=0.48\textwidth]{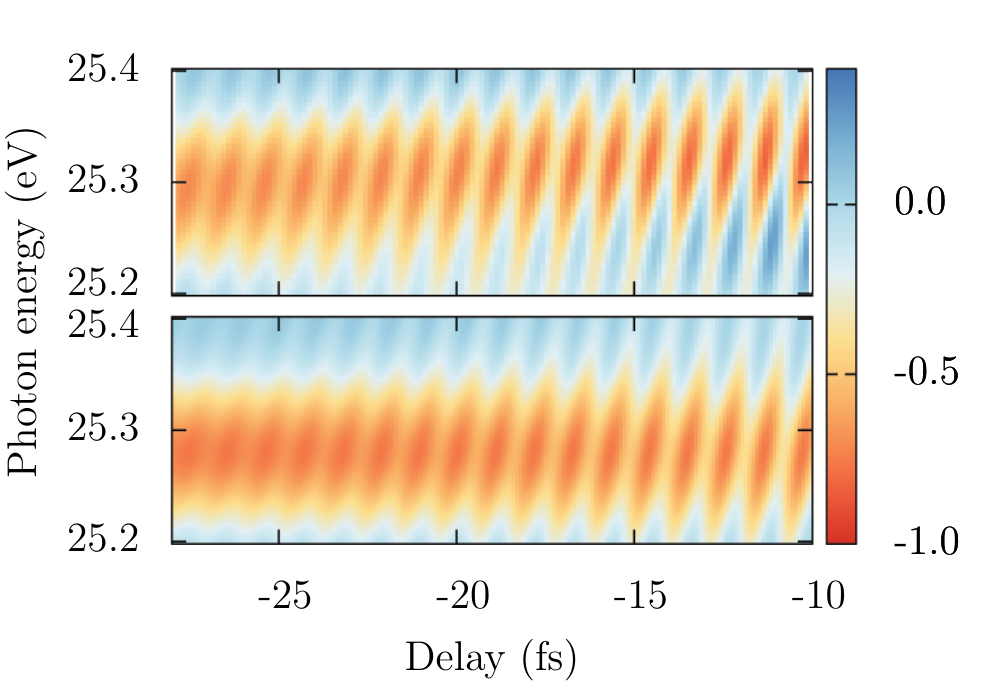}
\caption{\label{fringes} (Color online) Normalized ATAS spectrum $\tilde{S}(\omega,\tau)$ [Eq.~\eqref{response}] calculated for the fixed nuclei model of ${\text{H}_2}^+$. In the upper panel the Fourier transform of the time-dependent dipole moment is found from a full TDSE calculation [Eq.~\eqref{eq:TDSE_fixed}], while in the lower panel Eq.~\eqref{simple_dipole} was used. The pulse parameters are as in \fref{Fixed}.}
\end{figure}
In \fref{fringes} we compare the ATAS spectra constructed from dipole signals found from the full TDSE calculation [Eq.~\eqref{eq:TDSE_fixed}] and from Eq.~\eqref{simple_dipole}, respectively. Following Ref.~\citep{chen2013quantum} we have chosen $t_0$ to be located a quarter of a NIR period from the center of the NIR pulse. 
The amplitudes $c_\text{d}$ and $c_\text{id}$ can be found using perturbation theory, but the choice $c_\text{d}=c_\text{id}=0.5$ suffices to illustrate the predictions of the model.
As a result of the narrow spectral width of the NIR pulse, interference of the type seen in Eq. \eqref{simple_dipole} only occurs when the energy difference between two states of the same symmetry under $z \to -z$ is $\sim 2\omega_\text{NIR}$. Consequently the associated oscillation periods are always $\sim T_c^\text{NIR}/2$ with $T_c^\text{NIR}$ the period of a NIR pulse cycle.
We note that the interference phenomena of the type discussed here, always affect both absorption lines involved (at $E_4$ and $E_6$), and a similar pattern is found in the absorption line corresponding to the energy $E_4$. 
For this absorption line, however, the phase of Eq.~\eqref{simple_phi} is $(E_4-E_6)\times(t_0-\tau)$ and the tilt of the fringes in the interference pattern will be opposite to that seen in \fref{fringes}.
In Sec.~\ref{H2} we see that interference phenomena similar to the ones investigated here are also present between vibrational states in molecular systems, again leading to oscillations on the period of $T_c^\text{NIR}/2$.


%

\end{document}